\documentclass[11pt,a4paper]{article}
\pdfoutput=1

\usepackage{jheppub}
\usepackage[english]{babel}
\usepackage[utf8]{inputenc}
\usepackage{epsf}
\usepackage{graphicx}
\usepackage{subcaption} 

\usepackage{verbatim} 

\usepackage{amsmath}
\usepackage{amssymb}
\usepackage{mathrsfs}

\usepackage{mathbbol}

\usepackage{amsfonts}
\usepackage{bbding}
\usepackage{array}

\usepackage{color} 
\usepackage{tikz}
\usetikzlibrary{shapes,arrows,positioning,automata,backgrounds,calc,er,patterns}
\usepackage[compat=1.1.0]{tikz-feynman}
\vspace*{1cm}
\allowdisplaybreaks
\title{cLFV leptophilic $Z^\prime$ as a dark matter portal: prospects for colliders}

\author[1]{Andreas Goudelis}
\author[2]{\!\!, Jonathan Kriewald}
\author[1,3]{\!\!, Emanuelle Pinsard}
\author[1]{\!\!, Ana M. Teixeira}

\affiliation[1]{Laboratoire de Physique de Clermont (UMR 6533), CNRS/IN2P3, Univ.\ Clermont Auvergne, 4~Av.\ Blaise Pascal, F-63178 Aubi\`ere Cedex, France}
\affiliation[2]{Jožef Stefan Institut, Jamova Cesta 39, P. O. Box 3000, 1001 Ljubljana, Slovenia}
\affiliation[3]{Physik-Institut, Universität Zürich, CH-8057 Zürich, Switzerland}
\emailAdd{andreas.goudelis@clermont.in2p3.fr}
\emailAdd{jonathan.kriewald@ijs.si}\emailAdd{emanuelle.pinsard@physik.uzh.ch}
\emailAdd{ana.teixeira@clermont.in2p3.fr}

\abstract{
Extensions of the Standard Model featuring light vector bosons have been explored with the goal of resolving certain tensions between theory and experiment, among them the discrepancy in the anomalous magnetic moment of the muon, $\Delta a_{\mu}$. In particular, this is the case of a minimal construction including  a leptophilic, strictly flavour violating, vector boson $Z^\prime$. These new vector bosons are also well-motivated dark matter portals, with non-trivial couplings to stable, weakly interacting states which can account for the correct dark matter density. 
Here we study the prospects of a Standard Model extension  
(via a vector boson and a fermionic dark matter candidate)  concerning signatures at the LHC, and at future lepton and hadron colliders.
We discuss the cross-sections of several processes leading to same- and opposite-sign muon-tau lepton pairs in the final state, as well as final states with missing energy (in the form of neutrinos and/or dark matter).  
Our findings suggest that a future muon collider offers the best prospects to probe this model (together with searches for dilepton pairs and missing energy signatures at the FCC-ee running at the $Z$-pole); moreover, the complementarity of the different future high-energy colliders is also paramount to probing distinct $Z^\prime$ mass regimes.}

\begin{document}
\maketitle

\section{Introduction}\label{sec:intro}
The most recent result from the FNAL ``g-2'' E989 experiment~\cite{Muong-2:2023cdq} regarding the determination of the anomalous magnetic moment of the muon (in which data from all three runs have been combined), 
\begin{equation}\label{eq:exp-amu-2023}
a_\mu\, =\,  116 592 059(22) \times 10^{-11} \,,
\end{equation}
is in good agreement with the first disclosed result~\cite{Muong-2:2021ojo}, and represents a factor of two improvement in precision. When compared to the Standard Model (SM) theory prediction, 
$a_\mu^\text{SM}=116591810(43)\times 10^{-11}$ (as presented in the ``Theory initiative white paper''~\cite{Aoyama:2020ynm}), the new experimental result further hints at a strong discrepancy\footnote{Notice that here we follow~\cite{Aoyama:2020ynm}, and only consider the data-driven dispersive approach in the SM computation of the hadron vacuum polarisation contributions to $a_\mu^\text{SM}$, not taking into account the recent lattice QCD estimations.}, now standing at $5.0\sigma$.
Such a divergence from the SM prediction would necessarily call for New Physics (NP), already strongly needed in view of the SM's theoretical problems and of its observational caveats -- in particular neutrino oscillation data, the baryon asymmetry of the Universe and the need for a viable dark matter (DM) candidate. A number of well-motivated NP constructions, capable of addressing the tensions between experiment and the SM's prediction for the anomalous magnetic moment of the muon (i.e. $\Delta a_\mu$) have been investigated and extensively studied in recent years (for a review, see for instance~\cite{Athron:2021iuf}). 

Among the latter, extensions of the SM gauge group via additional $U(1)$ groups have been under intense scrutiny; not only can the associated new vector boson (possibly together with other new fields) ease certain tensions between the SM predictions and observation, but the new associated neutral currents can have a significant impact for particle and astroparticle physics~\cite{Langacker:2008yv}. In particular, such exotic vector bosons could act as a ``portal'' between the SM and exotic sectors containing viable dark matter candidates. Considering such a possibility could not only help us address the questions of the nature and abundance of DM in the Universe but, as we will also discuss in this paper, potential interactions of the new gauge bosons with the DM candidate could drastically alter their predicted phenomenology and provide new handles for their detection.

\medskip
Recently, the phenomenological impact of a minimal ad-hoc extension of the SM via leptophilic flavour violating 
$Z^\prime$ bosons was considered~\cite{Kriewald:2022erk}, aiming at assessing whether or not such an extension could simultaneously address the discrepancies in both $\Delta a_{e,\mu}$. Once all relevant constraints were taken into account
(in particular, electroweak precision observables, rare lepton flavour violating transitions and decays, as well as lepton flavour universality (LFU) tests in $Z$ and tau-lepton decays -- see also~\cite{Foot:1994vd,Heeck:2016xkh,Altmannshofer:2016brv,Buras:2021btx, Crivellin:2013hpa}), it was manifest that while such a minimal model could indeed saturate the tension in $\Delta a_{\mu}$, $\Delta a_{e,\tau}$ were predicted to be compatible with zero (in agreement with the SM's prediction). Preliminary studies of the possible signatures of such a lepton flavour violating (LFV) neutral vector boson at future colliders were also carried out. 

Given the promising initial studies of such an extension, and in view of the potential of a $Z^\prime$ boson to act as a portal to a ``dark'' sector, in what follows we extend the particle content of the minimal construction discussed in~\cite{Kriewald:2022erk} via the addition of a dark Majorana fermion, $\chi$, which only interacts with the SM through the new vector boson. As we will see, the presence of this new state, which can act as a DM candidate, also leads to new, unique decays (be it at the LHC, or at future lepton colliders). 

In the present study we thus address whether or not 
this minimal NP model can lead to ``observable'' events at high-energy colliders, under the requirement of explaining $\Delta a_\mu$, and providing a viable DM candidate compatible with existing DM searches.
Depending on the hierarchy of the masses of the $Z^\prime$ boson and of the DM candidate, one can expect a variety of collider signals, including opposite- and same-sign dilepton ($\mu-\tau$ pairs) final states, and possible missing energy signatures. In addition to the coming LHC runs (high-luminosity and possibly higher-energies), there are numerous proposals for a future generation of high-energy colliders: these include future circular colliders (such as FCC-ee and FCC-hh, or CEPC), linear colliders (as the ILC and CLIC) as well as muon colliders (including a future high-energy muon collider and other variants). Although we will not address the prospects   
for each facility, we will consider some representative examples, also discussing distinct operating modes (centre-of-mass energy and/or nature of the colliding beams). As subsequently mentioned upon the analysis, and in view of the unknown detector capabilities of most of the considered machines, we will only discuss the cross-sections associated with the channels that would allow to unveil the presence of this minimal NP model, and for certain cases, the expected number of events (however refraining from efficiency cuts, etc.).

Our study suggests that such a minimal SM extension
could indeed be at the source of a non-negligible number of events at high-energy colliders. In general, processes leading to same-sign $\mu\tau$ dilepton pairs (i.e. $\mu^\pm\mu^\pm \tau^\mp\tau^\mp$) would offer a true ``smoking gun'' for New Physics (in view of their absence in the SM). In general, and given the sizeable cross-sections and the observables to be studied, 
a future muon collider might be the optimal setup to probe and discover the NP model here analysed. Running at the $Z$-pole, FCC-ee also offers very promising prospects -- especially for opposite-sign $\mu\tau$ dilepton pairs in association with missing energy.

The manuscript is organised as follows: in Section~\ref{sec:model}, we describe the SM extension under study, 
discussing the distinct constraints on its parameter space (new couplings and masses of the NP states). In Section~\ref{sec:colliders}, we carry out a detailed analysis of the prospects of the present construction for current and future colliders (circular lepton and hadron, linear colliders and future muon colliders). The most important findings and a final discussion are summarised in the Conclusions. The Appendix contains complementary material (including relevant  constraints on flavour conserving and flavour violating processes, as well as illustrative topologies for the most relevant processes). 

\section{Flavour-violating leptophilic $Z^\prime$ and dark matter}\label{sec:model}
Let us begin by recalling the beyond-the-SM (BSM) construction proposed in~\cite{Kriewald:2022erk}, describing how it can also act as a portal to the DM sector. As already discussed in the Introduction, we extend the SM by an additional $Z^\prime$ gauge boson (which does not interact with the SM quarks), as well as by a SM gauge-singlet Majorana fermion $\chi$ only coupled to the $Z^\prime$ and which will act as our DM candidate.
Such an extension could be interpreted as the low-energy remnant of a complete ultraviolet (UV) construction which would include new fields responsible for the breaking of an additional $U(1)^\prime$ group (thus generating a mass for the new vector boson, and possibly for the  Majorana DM fermion state), among others. 

The interaction Lagrangian terms relevant for the present analysis are given by
\begin{eqnarray}\label{eq:lagrangian:Zchi}
  \mathcal{L}_{Z^\prime}^\text{int} &=&  \sum_{\alpha ,\beta} Z^\prime_\mu \left[ \bar{\ell }_\alpha \gamma^\mu \left(g_{X}^{\alpha \beta} \, P_X  \right)\ell_\beta + \bar{\nu}_\alpha \gamma^\mu \left(g_{L}^{\alpha \beta} \, P_L \right)\nu_\beta \right] + g_\chi \, Z^\prime_\mu \,\bar \chi \,\gamma^\mu \,\gamma^5\, \chi +
    \text{H.c.} \,,
\end{eqnarray}
in which the indices run over $\alpha ,\beta = e, \mu, \tau $, with $\alpha \neq \beta$; $P_X=P_{L,R}$ are the chiral projectors and $g_X^{\alpha \beta} = g_{L,R}^{\alpha \beta}$ are the new couplings (hermitian matrices in flavour space, that we will subsequently assume, for simplicity, to be real). The ``invisible'' coupling, $g_\chi$, is also taken to be real. 
We further notice that henceforth we will assume that both the $Z^\prime$ and the DM fermion $\chi$ are massive, but make no hypotheses regarding the origin of the masses, nor concerning the relative ordering of the  NP spectrum, i.e. $m_{Z^\prime} \gtrless m_\chi$. 
However, it is important to mention that we will always consider 
$m_{Z^\prime} \geq 10$~GeV, as a simple means to circumvent and evade numerous constraints from searches for such a new light vector at low energies (in particular indirect signals at $B$-factories~\cite{Iguro:2020rby}, or flavour violating $\tau\to\mu Z^\prime$ decays~\cite{Foot:1994vd,Heeck:2016xkh,Altmannshofer:2016brv}).

It should be noted that with the given particle content, and for generic choices of the couplings involved in the Lagrangian of Eq.~\eqref{eq:lagrangian:Zchi}, the model may predict non-vanishing gauge anomalies which would lead to a pathological behaviour of the theory in the ultraviolet. These anomalies can be cancelled in a variety of ways (for instance, through the introduction of new exotic fermions carrying non-trivial gauge charges and/or by invoking axions~\cite{Anastasopoulos:2006cz}). The impact of the precise anomaly-cancellation mechanism on low-energy observations is extremely difficult (if not impossible) to quantify in full generality. Given the fact that our analysis is purely driven by phenomenological considerations, throughout this work we will remain agnostic concerning this aspect of the theory's UV completion and treat all the couplings in Eq.~\eqref{eq:lagrangian:Zchi} as free parameters.

While the above leptophilic couplings of the $Z^\prime$ successfully allow the model to evade the constraints from meson oscillations and decays\footnote{New contributions to hadron transitions (induced from kinetic mixing) are negligibly tiny, and thus not included in the set of relevant constraints.} (arising from searches at NA48/2~\cite{Raggi:2015noa} and at KLOE-2~\cite{Anastasi:2015qla} experiments), which would be very stringent 
for a hadrophilic NP vector boson, 
flavour conserving lepton couplings (i.e. with a diagonal, albeit not necessarily universal structure, to both charged leptons and neutrinos) can still be at the source of new contributions to numerous observables, potentially in  disagreement with current bounds. 
Flavour conserving couplings to third and second generation leptons (including neutrinos) are strongly constrained by neutrino-electron scattering experiments, in particular from the data of the TEXONO~\cite{Deniz:2009mu} and CHARM-II~\cite{Vilain:1993kd} experiments, respectively for $\bar\nu_e$-electron and $\nu_\mu$-electron scattering. 
Neutrino trident production ($\nu_\mu N\to \nu_\mu N \mu^+\mu^-$)~\cite{CCFR:1991lpl} further constrains couplings to muons. In general, the most stringent bounds concern $Z^\prime ee$ couplings, as inferred from negative $Z^\prime$ searches at (electron) beam dump experiments 
(SLAC E141, Orsay, NA64~\cite{NA64:2018lsq, NA64:2019auh}), from dark photon production searches (KLOE-2 experiment~\cite{Anastasi:2015qla}, BaBar~\cite{BaBar:2014zli}) and finally from  parity-violation experiments (SLAC E158~\cite{SLACE158:2005uay}).

All the above referred bounds can be  circumvented for the choice of leptophilic, strictly flavour violating $Z^\prime$ couplings to leptons~\cite{Kriewald:2022erk}. Nevertheless, and as we proceed to discuss, numerous bounds still apply to the $g_{L,R}^{\alpha \beta}$-spanned parameter space, typically in association with charged LFV (cLFV) processes and searches for the violation of lepton flavour universality (LFUV). 

\subsection{Relevant experimental bounds on leptonic couplings}
Even if the off-diagonal structure of the new leptonic couplings allows evading numerous bounds, contributions to cLFV and LFUV-sensitive observables effectively constrain the allowed parameter space. This was extensively discussed in~\cite{Kriewald:2022erk}, to which we further refer for the detailed computation of the $Z^\prime$ contributions to the different observables. Below we briefly recall the  most relevant bounds\footnote{For completeness, let us notice that we do not have any significant NP contributions to $Z$ observables, such as the invisible $Z$ decay width or BR($Z\to \ell \ell$). } on each pair of couplings (under the assumption that the  other flavoured pairs have  been set to zero).  

\paragraph{$g_{L,R}^{e \mu }$ couplings}
The leading constraints arise from cLFV muonium-antimuonium oscillations (recalling that $\text{Mu}= \mu^+ e^-$); this can be readily understood since, and 
contrary to most cLFV models of NP, in the present construction
$\text{Mu}-\bar{\text{Mu}}$ oscillations occur at tree-level. 
The current bounds on this very rare process (see Appendix~\ref{app:flavour-bounds}) severely constrain both left- and right-handed couplings: for $m_{Z^\prime} = 10$~GeV, one finds $g_{L,R}^{e \mu} \lesssim 5  \times 10^{-3}$. It is also worth noticing that  LFUV-probes from ratios of $Z$ decays 
($R^Z_{\alpha\beta} = {\Gamma(Z \to \ell_\alpha^+\ell_\alpha^-)}/{\Gamma(Z \to \ell_\beta^+\ell_\beta^-)}$) also play a relevant, albeit subdominant role in constraining the 
$e-\mu$ couplings.

\paragraph{$g_{L,R}^{e \tau}$ couplings}
Precision searches for deviations from lepton flavour universality in $Z$ decays ($R^Z_{\alpha\beta}$) are also one of the most important constraints for the $e-\tau$ couplings. Equally important are LFUV probes in association with tau-lepton decays,  $R^\tau_{\mu e}$ $= {\Gamma(\tau^- \to \mu^- \nu_\tau \overline{\nu}_\mu)}/$ ${\Gamma(\tau^- \to e^- \nu_\tau \overline{\nu}_e)}$
(especially for regimes of $g_L^{e \tau} \approx 10^{-3}$). 

\paragraph{$g_{L,R}^{\mu \tau}$ couplings}
Finally, special attention must be devoted to the $\mu-\tau$
couplings, especially since these are at the origin of the contributions required to saturate the discrepancy in 
$(g-2)_\mu$, as we will discuss in the following subsection.
In this case, the leading constraints also arise from both 
$R^Z_{\alpha\beta}$ and $R^\tau_{\mu e}$.

\subsection{Explaining the anomalous magnetic moment of the muon}
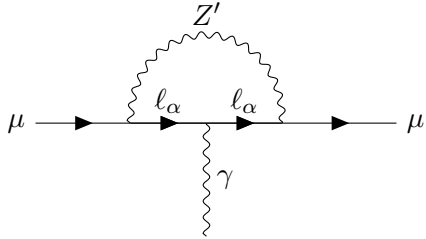
\begin{figure}[h!]
    \centering
\begin{tikzpicture}
\begin{feynman}
\vertex (a1) {\(\mu\)};                    
\vertex[right=1.5cm of a1] (a2);
\vertex[right=1cm of a2] (a3);
\vertex[right=1cm of a3] (a4);
\vertex[right=1.5cm of a4] (a5) {\(\mu\)};
\vertex[below=1.5cm of a3] (b);
\diagram* {
{[edges=fermion]
(a1) -- (a2) -- [fermion, edge label=\(\ell_\alpha\)](a3) -- [fermion, edge label=\(\ell_\alpha\)](a4) -- (a5),
},
(a3) -- [boson, edge label=\(\gamma\)] (b),
(a4) -- [boson, out=90, in=90, looseness=2.0, edge label'=\(Z^\prime\)] (a2)
};
\end{feynman}
\end{tikzpicture}
    \caption{New contributions to the anomalous magnetic moment of the muon (notice that the internal lepton flavour is necessarily different from the external one, $\ell_\alpha \neq \mu$).}
    \label{fig:aell}
\end{figure}

As discussed in detail in~\cite{Kriewald:2022erk}, the anomalous magnetic moment of the muon receives new contributions from the $Z^\prime$ boson at the 1-loop level, owing to the LFV interactions with the external muon lines, see Fig.~\ref{fig:aell} (we remind that the internal lepton is an $e$ or a $\tau$). 

\noindent 
Following~\cite{Lindner:2016bgg, Jegerlehner:2009ry, Leveille:1977rc}, the  $Z^\prime$-induced contributions to 
$\Delta a_\mu$ are given by 
\begin{equation}
    \Delta a_\mu \,= \,\sum_i \left[\dfrac{|g_V^{\mu \alpha}|^2}{4\pi^2}\,\dfrac{m_\alpha^2}{m_{Z^\prime}^2}\,F(\lambda, \epsilon_\alpha) + \dfrac{|g_A^{\mu \alpha}|^2}{4\pi^2}\,\dfrac{m_\mu^2}{m_{Z^\prime}^2}\,F(\lambda, -\epsilon_\alpha)\right]\,.
\label{eq:delta-amu-NP}
\end{equation}
In the above equation we have introduced the convenient notation
$g_V, g_A = (g_L \pm g_R)/2$ (corresponding to the  
vector and axial-vector couplings, respectively); 
the sum runs over the internal leptons and the loop function $F(\lambda, \epsilon_\alpha)$ can be expressed as 
\begin{align}
       F(\lambda, \epsilon_\alpha) &= \dfrac{1}{2}\int^1_0 dx \left[ \dfrac{2x(1-x)[x-2(1-\epsilon_\alpha)] }{(1-x)(1-\lambda^2x)+\epsilon_\alpha^2\lambda^2x} + \dfrac{\lambda^2x^2(1-\epsilon_\alpha)^2(1+\epsilon_\alpha-x)}{(1-x)(1-\lambda^2x)+\epsilon_\alpha^2\lambda^2x}\right] \,, 
\label{eq::functionfordeltaa}
\end{align}
with $\epsilon_\alpha = m_\alpha/m_\mu$, in which $m_\alpha$ 
is the internal fermion mass and $\lambda = m_\mu/m_{Z^\prime}$.
From Eq.~(\ref{eq:delta-amu-NP}), it is clear that the most significant NP contributions to $\Delta a_\mu$ will be  produced in association with the propagation of a tau lepton in the loop, and thus be proportional to $(g_{L,R}^{\mu \tau})^2$.
According to the findings of~\cite{Kriewald:2022erk}, and for $m_{Z^\prime} =10$~GeV, the $\mathcal{O}(5\sigma)$ tension on $(g-2)_\mu$ can  be accommodated for $g_{L}^{\mu \tau} \sim \mathcal{O}(10^{-3})$, and $g_{R}^{\mu \tau} \sim \mathcal{O}(10^{-2})$, while in agreement with all experimental constraints (which stem from LFUV in $Z$ and $\tau$ decays).

However, it is important to emphasise that once non-vanishing values for $g_{L,R}^{\mu \tau}$ are taken into account, the 
previously  derived bounds on $g_{L,R}^{e \tau}$ and $g_{L,R}^{e \mu}$ couplings will be also impacted, in view of the new contributions to several observables (potentially in conflict with experimental measurements and bounds).
As discussed in~\cite{Kriewald:2022erk} for $g_{L,R}^{e \mu}$, muonium oscillation bounds are now clearly superseded by rare cLFV tau decays, in particular $\tau \to e \bar \mu \mu$ and 
$\tau \to \mu \bar e \mu$, imposing
$g_{L,R}^{e \mu} \lesssim 2 \times 10^{-5}$; likewise, and for 
$g_{L,R}^{e \tau}$, the former leading constraints from LFUV observables $R^Z_{\alpha \beta}$ and $R^\tau_{\mu e }$ become subleading, with the most constraining observables now being cLFV muon decays (neutrinoless $\mu-e$ conversion, 3-body and radiative muon decays), with $\mu \to e \gamma$ forcing 
$g_{L,R}^{e \mu} \lesssim \mathcal{O}( 10^{-7})$. 
For heavier mass regimes, the above limits on the cLFV couplings of the $Z^\prime$ are relaxed (see~\cite{Kriewald:2022erk}).

In order to summarise the above discussion, in Table~\ref{tab:couplings_masses}
we collect information on (best fit) values\footnote{We notice that should the tension in $(g-2)_\mu$ be reduced, one would be led to regimes of smaller $g_{L,R}^{\mu\tau}$ couplings.} and associated uncertainties of the $\mu-\tau$ couplings for a set of benchmark $Z^\prime$ masses (assuming vanishing values of the $g_{L,R}^{e \ell}$ couplings), which will be used throughout the subsequent dark matter and collider studies. 
\renewcommand{\arraystretch}{1.3}
\begin{table}[h!]
    \centering
    \hspace*{-7mm}{\small\begin{tabular}{|c|c|c|}
    \hline
    $m_{Z^\prime}$ (GeV) &$g_L^{\mu\tau}$ & $g_R^{\mu\tau}$  \\
    \hline\hline
    $10$ & \: $0.0024 \pm 0.0005$ \: & \: $0.036 \pm 0.013 $ \:  \\
    \hline
    $ 20$ & $0.0045 \pm 0.001$   & $ 0.073 \pm 0.028$  \\
    \hline
    $ 30 $ & $0.0067  \pm 0.0015$   & $0.11 \pm 0.04$  \\
    \hline
    $ 50$ & $0.0113  \pm 0.0026$   & $ 0.17 \pm 0.07$  \\
    \hline
    $ 100$ & $ 0.024 \pm 0.006$   & $0.3  \pm 0.1$  \\
    \hline
    $ 150$ & $ 0.037 \pm 0.008$   & $ 0.41 \pm 0.13$  \\
    \hline
    $ 200 $ & $ 0.050 \pm 0.011$   & $0.53 \pm  0.16$  \\
    \hline
    \end{tabular}}
    \caption{Best fit values and associated uncertainties of the $\mu-\tau$ couplings for different $Z^\prime$ masses; the fit includes $(g-2)_\mu, \, R^Z_{\alpha\beta}$ and $R^\tau_{\mu e}$ constraints (adapted and updated from~\cite{Kriewald:2022erk}).}
    \label{tab:couplings_masses}
\end{table}
\renewcommand{\arraystretch}{1.}

\subsection{Model implementation for numerical studies}
In order to efficiently perform our numerical analyses for dark matter and collider phenomenology of the minimal setup described by the Lagrangian of Eq.~\eqref{eq:lagrangian:Zchi}, we have implemented the model in the {\tt FeynRules} package~\cite{Alloul:2013bka}. Here we briefly provide some short explanations concerning our implementation. 

Recall that in essence, the model contains two new states:
the exotic vector $Z^\prime$, labelled {\tt Zp} and
the (Majorana) dark fermion $\chi$, labelled {\tt chi}. 
All in all, the model introduces five new parameters: the $Z^\prime$ and dark matter masses, labelled {\tt MZp} and {\tt mchi} respectively;
the flavour-violating $Z^\prime$ left- and right-handed couplings, labelled {\tt gmtl} and {\tt gmtr} respectively and
the $Z^\prime$ coupling to dark matter, called {\tt gchi}.
With this implementation at hand, we now proceed to compute the relevant observables and discuss our results.

We notice that the relevant model files will be uploaded to the {\tt FeynRules} Model Database in the near future.

\subsection{Dark matter relic density}
The observed~\cite{Planck:2018vyg} dark matter relic abundance
\begin{equation}\label{eq:Planck2018}
    \Omega_{\rm DM} h^2 \,=\, 0.12 \pm 0.0012\,,
\end{equation}
can be reproduced, in our model, in the context of (at least) two distinct thermodynamical settings. If the interactions between the $Z^\prime$ and the dark matter candidate $\chi$ are strong enough, at sufficiently high cosmic temperatures the latter will equilibrate with the SM thermal bath\footnote{Note that the sizeable interactions between the $Z^\prime$ and the SM particles required by the flavour considerations presented in the previous sections also imply that the $Z^\prime$ is itself part of the SM thermal bath.}. In this case, the dark matter relic abundance is predicted by the usual thermal freeze-out mechanism. In the other extreme, that is if the $Z^\prime \chi\chi$ coupling is very small, $\chi$ will never reach thermal equilibrium with the bath particles, and in this case we should rather place ourselves within the context of the freeze-in mechanism. Even though from a cosmological standpoint both scenarios are equally viable, from a collider perspective the detection of a feebly interacting massive particle
(FIMP)-like dark matter candidate such as the one considered in this work is extremely challenging, since if $\chi$ is feebly interacting it cannot be produced at colliders at observable rates. In the remainder of this study we will, hence, focus on the freeze-out case.

\begin{figure}[h!]
    \centering
    \includegraphics[width=.6\textwidth]{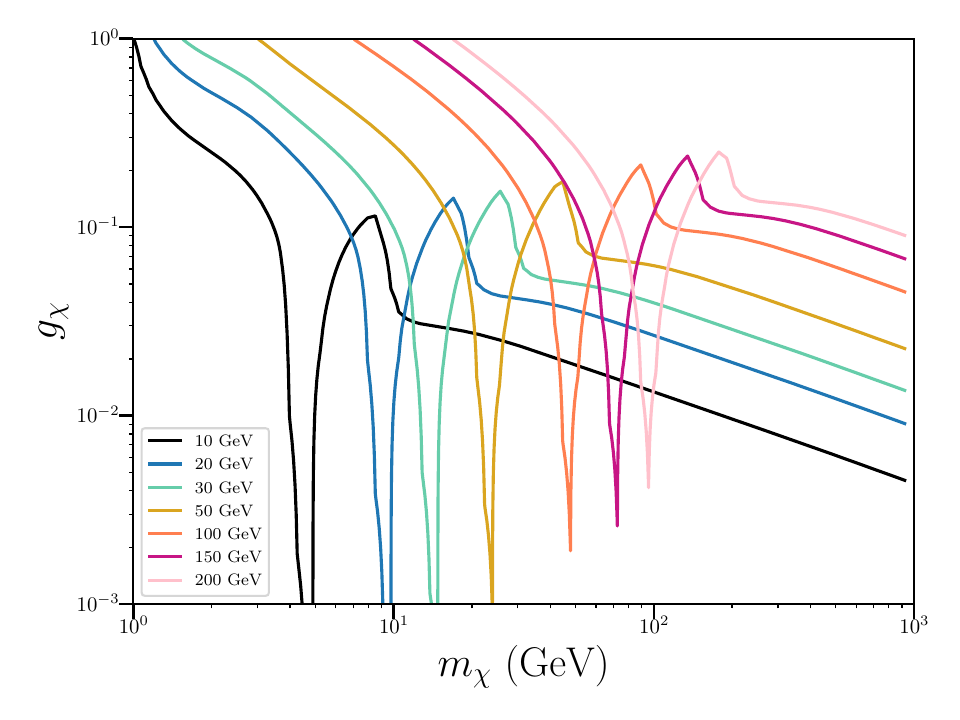} \hfill
    \caption{Contours of $(m_\chi, g_\chi)$ combinations for which the Planck 2018 constraint of Eq.~\eqref{eq:Planck2018} is satisfied within $3\sigma$ from the central value for the seven $m_{Z^\prime}$ benchmarks presented in Table~\ref{tab:couplings_masses}.}
    \label{fig:omega}
\end{figure}

In our model thermal freeze-out proceeds through annihilation of $\chi$ pairs into $\nu \bar{\nu}$, $\tau^\pm \mu^\mp$ and/or $Z^\prime Z^\prime$ pairs, depending on whether or not the latter two channels are kinematically accessible. When all annihilation channels are open, the leading contribution is dictated by the hierarchy between the couplings $g_{L/R}^{\alpha \beta}$ and $g_\chi$, where by $g_{L/R}^{\alpha \beta}$ we collectively denote the couplings of the $Z^\prime$ to the SM leptons. 
In order to compute the predicted dark matter relic abundance we employ the {\tt micrOMEGAs 5} numerical code~\cite{Belanger:2018ccd}, upon generation of the necessary {\tt CalcHEP}~\cite{Belyaev:2012qa} model files through our {\tt FeynRules} implementation previously described. Our results are presented in Fig.~\ref{fig:omega}, in which we vary $(m_\chi, g_\chi)$ and compute the predicted DM abundance for the seven $m_{Z^\prime}$ benchmark scenarios defined in Table~\ref{tab:couplings_masses}.

For all scenarios, the cosmologically viable parameter space exhibits a behaviour which is fairly typical of minimal $Z^\prime$-mediated dark matter models. For each benchmark, when the $\chi$ mass is low, $m_\chi \ll m_{Z^\prime}/2$, dark matter is mainly depleted through $s$-channel $Z^\prime$-mediated annihilation into $\tau^\pm \mu^\mp$ pairs\footnote{Notice that, since for the benchmarks presented in Table~\ref{tab:couplings_masses} $g_{L}^{\mu\tau}$ is consistently smaller than $g_{R}^{\mu\tau}$, annihilation into neutrino pairs turns out to be subleading as long as annihilation into $\tau^\pm \mu^\mp$ pairs is kinematically allowed.} with the relevant cross-section scaling roughly as $\left(g^{\alpha \beta}\, g_{\chi}/m_{Z^\prime}^2\right)^2$. In these regions of the parameter space, the required $g_\chi$ coupling values are fairly large, reaching ${\cal{O}}(1)$ or even non-perturbative values when $m_\chi \ll m_{Z^\prime}$. As the limit $m_\chi \sim m_{Z^\prime}/2$ is approached from below, the annihilation cross-section is significantly (resonantly) enhanced and the observed dark matter abundance is obtained for much lower dark coupling ($g_{\chi}$) values. As the dark matter mass increases further, the resonant enhancement effect becomes attenuated and the required value of the coupling increases again, until $t$-channel $\chi$-mediated annihilation into $Z^\prime Z^\prime$ pairs becomes kinematically accessible. Once this happens, the total dark matter annihilation cross-section is boosted and the required values of $g_\chi$ decrease once more by a factor of a few. Note that, in this regime, the annihilation mode $\chi \chi \rightarrow Z^\prime Z^\prime$ generically dominates the total DM self-annihilation cross-section, and scales roughly as $g_\chi^4 m_\chi^2/m_{Z^\prime}^2$ (see for instance,~\cite{Arcadi:2017kky}). This behaviour explains the fact that, in the $\chi$ mass range in which all annihilation channels become available, the required coupling values tend to decrease with increasing $m_\chi$. Let us also notice that the $1/m_{Z^\prime}^2$ suppression of $\left\langle \sigma v \right\rangle$ is the reason why, for increasing values of the $Z^\prime$ mass (and for fixed values of the $Z^\prime$ couplings to the SM), a larger $g_\chi$ is needed in order to match the Planck measurements. 

The presence of the dark state $\chi$ can drastically modify the $Z^\prime$ collider phenomenology, depending on one hand on whether or not the decay $Z^\prime \rightarrow \chi \chi$ is allowed and, on the other hand, on the hierarchy between the dark and visible couplings of the $Z^\prime$ (the latter being dictated through the interplay of the requirements to explain $(g-2)_\mu$ and the Planck observations). In this spirit, in Table~\ref{tab:colliderbenchmarks} we present a set of benchmarks whose phenomenology will be analysed in the following sections. In particular, for each of the benchmark points presented in Table~\ref{tab:couplings_masses} we now distinguish two different scenarios satisfying the relic density constraint of Eq.~\eqref{eq:Planck2018}: one in which the decay $Z^\prime \rightarrow \chi\chi$ is kinematically forbidden, and another in which it is allowed.

When necessary to identify specific benchmark points, we will use the following labelling scheme: $Z^\prime_{X}\chi_{Y}$, in which $X,Y$ will be replaced by the corresponding values of 
$m_{Z^\prime}$ and  $m_\chi$ (in GeV). For example, the last line of Table~\ref{tab:colliderbenchmarks} would correspond to 
$Z^\prime_{200}\chi_{300}$.
\renewcommand{\arraystretch}{1.3}
\begin{table}[h!]
    \centering
    \hspace*{-7mm}{\small\begin{tabular}{|c|c|c|}
    \hline
    $m_{Z^\prime}$ (GeV) & $m_\chi$ (GeV) & $g_\chi$ \\
    \hline\hline
    $10$ & $2$ & $0.25$ \\
    \hline
    $10$ & $10$   & $0.04$ \\
    \hline
    $20$ & $5$   & $0.178$ \\
    \hline
    $20$ & $40$   & $0.039$ \\
    \hline
    $30$ & $10$   & $0.098$ \\
    \hline
    $30$ & $50$   & $0.050$ \\
    \hline
    $50$ & $10$   & $0.275$ \\
    \hline
    $50$ & $60$   & $0.070$ \\
    \hline
    $100$ & $20$   & $0.32$ \\
    \hline
    $100$ & $200$   & $0.089$ \\
    \hline
    $150$ & $20$   & $0.578$ \\
    \hline
    $150$ & $200$   & $0.118$ \\
    \hline
    $200$ & $50$   & $0.260$ \\
    \hline
    $200$ & $300$   & $0.134$ \\
    \hline
    \end{tabular}}
    \caption{Benchmark scenarios based on the results presented in Table~\ref{tab:couplings_masses} for which the dark matter relic density constraint is satisfied and which are expected to exhibit distinct phenomenological features at different colliders.}
    \label{tab:colliderbenchmarks}
\end{table}
\renewcommand{\arraystretch}{1.}

\subsection{Constraints from dark matter searches}
Given the fact that throughout our analysis we are only assuming (off-diagonal) $Z^\prime$ couplings to second and third generation leptons, our model does not predict visible signals in direct detection experiments. 

However, indirect searches for visible dark matter annihilation products in the galaxy and beyond, most notably gamma-rays, could be sensitive to some part of the parameter space. The strongest constraints from gamma-ray searches have been set through the combined analysis of MAGIC and Fermi-LAT observations of dwarf spheroidal galaxies~\cite{MAGIC:2016xys}, which where reanalysed in~\cite{Alvarez:2020cmw} using data-driven astrophysical $J$-factors. Both analyses present their results in terms of a set of specific dark matter annihilation final states. For a $10$ GeV dark matter mass, the experimental collaborations reach a sensitivity of the order of $10^{-26}$~cm$^3$/s assuming annihilation into $\tau^+ \tau^-$ pairs, or $10^{-25}$~cm$^3$/s assuming annihilation into $\mu^+ \mu^-$. Similar results were presented in the latter reanalysis, whereas in~\cite{Fermi-LAT:2013sme} these limits are shown to improve by a factor of a few as the dark matter mass extends down to $\sim 1$~GeV.

Moreover, dark matter annihilation could also affect the Cosmic Microwave background (CMB): the energy injected in the plasma may change the ionisation history of the Universe which, in turn, can alter the CMB angular power spectrum. The analysis performed in~\cite{Kawasaki:2021etm} estimated the corresponding limit on the DM self-annihilation cross-section to be of the order of $2\times 10^{-27}$ cm$^3$/s for a 1~GeV weakly interacting massive particle (WIMP) annihilating into muon pairs, whereas above $10$~GeV the limit becomes subleading with respect to the MAGIC/Fermi-LAT one.

Rigorously applying these limits in our model requires a dedicated analysis that falls beyond the scope of the present paper, given that the dark matter annihilation final states considered in these works do not match the ones present in our framework. We have, however, computed the channel-by-channel zero-velocity dark matter self-annihilation cross-section, relevant both for indirect detection and for CMB observations, and we will limit ourselves to a few remarks concerning our findings. Concretely, the low regime of our lowest-mass $m_{Z^\prime}$ benchmark ($m_\chi < 4$~GeV) could indeed be in some tension with the CMB observations, given that we find it to predict a DM annihilation cross-section of the order of $2\times 10^{-26}$~cm$^3$/s for $m_\chi = 1$~GeV, falling rapidly below $10^{-27}$~cm$^3$/s as the DM mass approaches $4$~GeV (i.e. as the $Z^\prime$ resonance is approached). As the DM mass increases further, $\left\langle \sigma v\right\rangle$ once more reaches the value of $10^{-26}$~cm$^3$/s for $m_\chi \sim 10$~GeV, only to decrease rapidly again for higher masses (i.e. once the $Z^\prime Z^\prime$ final state opens up). Similar remarks apply to the $m_{Z^\prime} = 20$~GeV scenario, whereas for higher $Z^\prime$ masses the zero-velocity self-annihilation cross-section consistently lies well below the indirect detection/CMB limits on $\left\langle \sigma v\right\rangle$.

\subsection{Unitarity constraints}
As already mentioned, the model described by the Lagrangian in Eq.~\eqref{eq:lagrangian:Zchi} is by no means intended to serve as a UV-complete extension of the Standard Model of particle physics. In particular, the $Z^\prime$ vector boson mass is generically expected to stem from some Higgs-like breaking of an extension of the SM gauge group. In the spirit of remaining as general as possible, throughout the analysis performed in this paper we treat $m_{Z^\prime}$ as a free parameter. However, despite the fact that the concrete structure of potential SM extensions (which could amount to a low-energy Lagrangian of the form of Eq.~\eqref{eq:lagrangian:Zchi}) is extremely diversified, it is important to keep track of a set of minimal theoretical requirements ensuring that our predictions hold -- at least at a qualitative level -- for a reasonable set of models.

Concretely, it is known that extensions of the SM involving new vector bosons, most notably with axial couplings to fermions (be them SM fermions or exotic ones), can challenge high-$\sqrt{s}$ scattering matrix unitarity considerations (for a recent reference see e.g.~\cite{Kahlhoefer:2015bea}). These constraints inevitably do involve some degree of arbitrariness; indeed, although one can define quantitatively specific unitarity-breaking criteria, the latter
\begin{itemize}
    \item can vary depending on concrete numerical considerations (i.e. the exact numerical value at which unitarity of the $S$ matrix is taken to be violated).
    \item are extremely difficult -- if not impossible -- to model in full generality, since the way through which different sectors work together to preserve unitarity is highly \textit{model-dependent}.
\end{itemize}
In our case, a potential bound could be related to the scattering amplitude of the process $\chi \chi \rightarrow Z^\prime Z^\prime$, which is highly relevant for dark matter annihilation. For each $(m_{Z^\prime}, m_\chi, g_\chi)$ combination, this process leads to an upper bound on the centre-of-mass energies that can be considered \cite{Kahlhoefer:2015bea}
    \begin{equation}
        \sqrt{s} < \frac{\pi m_{Z^\prime}^2}{(g_\chi)^2 m_\chi}\,.
    \end{equation}
In the case of dark matter annihilation, we can further approximate $\sqrt{s} \approx 2m_\chi$ (since, during freeze-out, the kinetic energy of DM particles for the mass range we consider is negligible compared to their mass) and, rather convert this limit to a bound on the dark matter mass
    \begin{equation}\label{eq:unitarityconstraint}
        m_\chi < \sqrt{\frac{\pi}{2}} \frac{m_{Z^\prime}}{g_\chi}\,.
    \end{equation}
It turns out that, for the benchmark $m_{Z^\prime}$ values we consider, the $(m_\chi, g_\chi)$ combinations complying with the Planck bounds consistently satisfy this unitarity constraint.

\section{Collider prospects}\label{sec:colliders}

\renewcommand{\arraystretch}{1.3}
\begin{table}
    \centering
    \begin{tabular}{|c|c|c|}
        \hline 
        Facility & $\sqrt{s}$ & Integrated luminosity (total)\\
        \hline \hline
        LHC  & 13 TeV & 190 fb$^{-1}$ - plan 450 fb$^{-1}$ \\
        \hline HL-LHC~\cite{Apollinari:2017lan} & 13.6 TeV & 3000-4000 fb$^{-1}$ \\
        \hline HE-LHC~\cite{FCC:2018bvk} & 27 TeV & 10 ab$^{-1}$ \\
        \hline FCC-hh~\cite{FCC:2018vvp}& 100 TeV & 20 ab$^{-1}$ (combine to 30 ab$^{-1}$) \\
        \hline
    CLIC~\cite{CLIC:2016zwp} & 380 GeV & 500 fb$^{-1}$~\cite{Aicheler:2012bya}\\
        & 1.5 TeV & 1500 fb$^{-1}$  \\
     & 3 TeV  & 3000 fb$^{-1}$ \\
    \hline
    FCC-ee~\cite{FCC:2018evy}& 91 GeV & 192 ab$^{-1}$ \\ 
    & 161 GeV & 12 ab$^{-1}$\\ 
    & 240 GeV & 5.1 ab$^{-1}$\\ 
    & 350-365 GeV & 1.7 ab$^{-1}$\\ 
    \hline
    Muon collider~\cite{MuonCollider:2022nsa,MuonCollider:2022xlm} & 3 TeV & 1 ab$^{-1}$ \\
        & 10 TeV & 10 ab$^{-1}$\\
    \hline
    $\mu$TRISTAN~\cite{Hamada:2022mua}: $\mu^+\mu^+$ & 2 TeV &  assume 120 fb$^{-1}$~\cite{Hamada:2022uyn}\\
    $\mu$TRISTAN: $\mu^+e^-$ & 346 GeV (30 GeV $e$ + 1 TeV $\mu$ beam) & 1 ab$^{-1}$ \\
    \hline
    \end{tabular}
    \caption{Expected centre-of-mass energy and integrated luminosity at different colliders. }
    \label{tab:facility_energies}
\end{table}
\renewcommand{\arraystretch}{1}

Having presented the relevant constraints on the parameter space (spanned by the new mediator's mass and associated couplings), we now proceed to assess the prospects for collider searches of such a SM extension via a cLFV leptophilic $Z^\prime$ and a Majorana 
dark matter candidate $\chi$. In what follows we will discuss different potential experimental settings and detection channels, especially focusing on final states with little or no SM background. 

Concretely, on the side of experimental facilities we will consider $pp$ collisions at the LHC, $e^+ e^-$ collisions at a future FCC-ee collider (running at the $Z$ pole and at the $WW$, $HZ$ and $t\bar t$ thresholds) and, finally, two possible muon machines -- a generic muon collider and $\mu$TRISTAN. For completeness we will also include the prospects for an FCC-hh machine at its peak energy runs (close to 100~TeV). The different projected centre-of-mass energy and luminosity configurations of such facilities are collected in Table~\ref{tab:facility_energies}.

As far as experimental signatures are concerned, we will study the following processes\footnote{It is useful to clarify the notation  concerning the collider analysis. Firstly, in an experimental setting $\tau$-leptons essentially behave as jets. Nonetheless, in the remainder of this paper, we will adopt a ``theory-driven'' scheme
and refer to $\tau$'s as leptons. Secondly, we will refer to final states of the type $\mu^-\mu^+\tau^-\tau^+$ as ``opposite-sign'' dileptons, whereas final states of the type $\mu^-\mu^-\tau^+\tau^+$ will be referred to as ``same-sign'' dileptons.}
\begin{eqnarray}
    f\bar{f} \to \mu^-\mu^+\tau^-\tau^+\,, \quad \quad f\bar{f} \to \mu^-\mu^-\tau^+\tau^+\,,  \quad \quad f\bar{f} \to \mu^-\tau^+ + E^\mathrm{miss}\,, 
\end{eqnarray}
in which $f\bar{f}$ corresponds to $q\bar{q}, \, e^+e^-$ or $\mu^+\mu^-$, depending on whether the corresponding machine is a hadron, $e^+e^-$ or $\mu^+\mu^-$ collider; the missing energy (transverse, in the case of hadron colliders)  is due to the presence of neutrinos and/or DM particles in the final state, i.e. $E^\mathrm{miss}= \bar\nu_\mu\nu_\tau,\, \nu_\mu\bar\nu_\tau, \,\chi \chi $. 
Moreover, for muon colliders we will also consider the processes $\mu^- \mu^+\to \tau^-\tau^+  + E^\mathrm{miss}$; in this case, $E^\mathrm{miss}= \bar\nu_\mu\nu_\mu,\, \nu_\tau\bar\nu_\tau, \,\chi \chi $.
In addition, and concerning $\mu$TRISTAN, we will further study the following signatures
\begin{align}
    \mu^+\mu^+  &\to \mu^+\mu^-\tau^+\tau^+\,,  &\mu^+\mu^+ &\to \mu^+\mu^+\tau^-\tau^+\,, \nonumber\\
     \mu^+\mu^+ &\to \mu^+\tau^+ + E^\mathrm{miss}\,,  &\mu^+\mu^+ &\to \tau^+\tau^+ + E^\mathrm{miss}\,, \nonumber \\
     e^-\mu^+ &\to e^-\mu^-\tau^+\tau^+\,,  &e^-\mu^+ &\to e^-\mu^+\tau^-\tau^+\,, \quad \quad 
     e^-\mu^+ \to e^-\tau^+ + E^\mathrm{miss}\,. 
\end{align}
We notice that in all the processes above mentioned we do not consider Higgs-mediated contributions.

\paragraph{Event simulation}
Once the relevant signal processes and background contributions 
have been identified, we generate the events using {\tt FeynRules/MadGraph5\_aMC@NLO}~\cite{Alwall:2014hca}, for the aforementioned different collider scenarios (and when relevant, for distinct values of $\sqrt s$).  Moreover, in the following we use the cuts proposed in~\cite{Altmannshofer:2016brv} for processes leading to 4 visible leptons in the final state, and for final states with two leptons and two invisible particles (neutrinos and/or DM): leading lepton transverse momentum $p_T>20$ GeV, sub-leading lepton(s) $p_T>15$ GeV, lepton pseudo-rapidity $|\eta|<2.7$ and lepton isolation $\Delta R>0.1$.
As mentioned before, we notice that in view of the uncertainty concerning future collider projects -- or even upgrades of existing colliders -- in the present study we refrain from carrying out any detector simulation. 

\subsection{Hadron ($pp$) colliders}
\begin{figure}[h!]
    \centering
\begin{tikzpicture}
    \begin{feynman}
    \vertex (q1) at (-1,1) {\(f\)};
    \vertex (q2) at (-1,-1){\(\bar f\)};
    \vertex (a) at (-0 ,0) ;
    \vertex (b) at (2,0);
    \vertex (c) at (4,1.4) {\(\tau^+/\mu^+\)};
    \vertex (d) at (4,-1.4){\(\mu^-/\tau^-\)}; 
    \vertex (e) at (3,0.7);
    \vertex (f) at (3.7,0);
    \vertex (f1) at (5.3,0.7){\small\(\mu^+ / \tau^+ /\bar \nu / \bar\chi \)};
    \vertex (f2) at (5.3,-0.7){\small\(\tau^- /\mu^- / \nu / \chi \)};
    \diagram* {
    (q1) -- [fermion] (a) -- [fermion] (q2),
    (a) -- [boson, edge label=\(Z\gamma\)] (b),
    (c) -- [fermion] (e) -- [fermion, edge label'=\small{\(\mu^+/\tau^+\)}] (b) -- [fermion] (d),
    (e) -- [boson, edge label=\(Z^\prime\)] (f),
    (f1) -- [fermion](f) -- [fermion](f2)};
    \end{feynman}
\end{tikzpicture}
    \caption{Illustrative example of a possible topology for the considered processes at hadron (and also electron) colliders.}
    \label{fig:ffcollisions}
\end{figure}
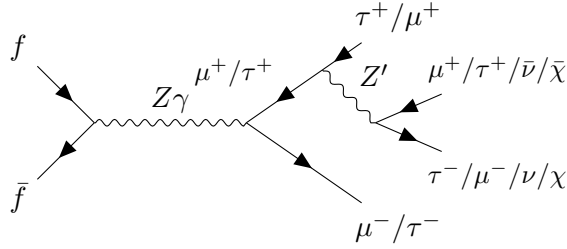

We begin our numerical analysis with the prospects for the existing/planned LHC runs (assuming, for simplicity, a 14~TeV centre-of-mass energy), as well as those of a future hadron collider (which we illustrate by means of the proposed FCC-hh).
Figure~\ref{fig:ffcollisions} illustrates a possible topology for the processes we proceed to discuss. Interestingly, this is the only topology contributing to both $e^+e^-$ and $pp$ colliders, at the origin of same-sign dilepton pairs. Concerning opposite-sign dilepton pairs and  final states with missing energy, this is also the only topology to which the NP states contribute.

\paragraph{LHC runs at 14 TeV}

\begin{figure}[h!]
    \centering
    \includegraphics[width=0.49\textwidth]{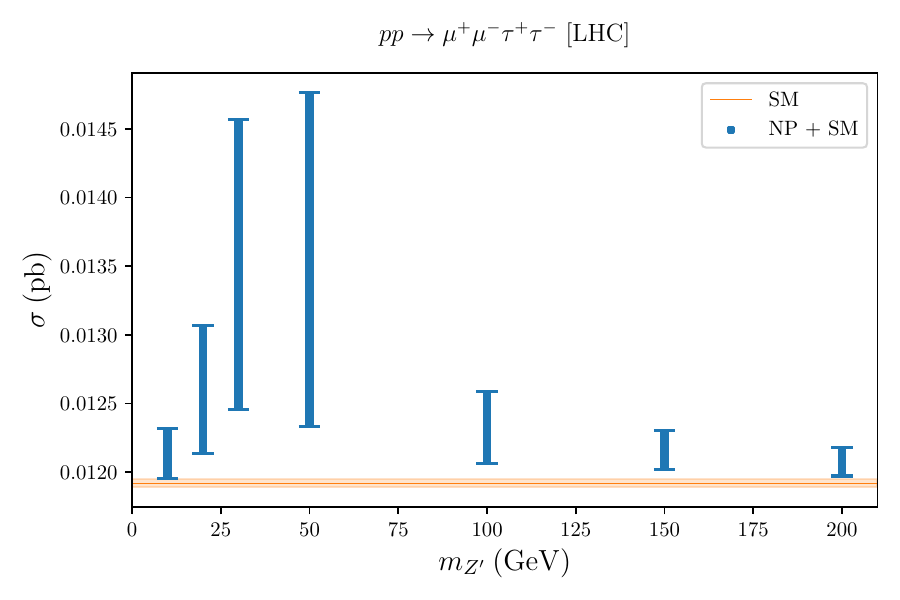} \hfill
    \includegraphics[width=0.49\textwidth]{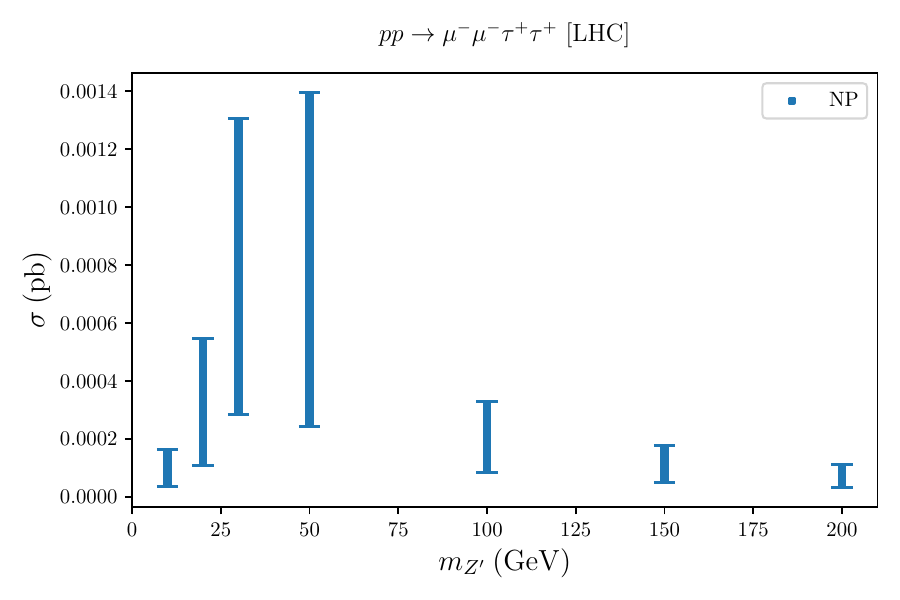} 
    \caption{LHC: prospects for the cross-sections, $\sigma(pp\to\mu^+\mu^-\tau^+\tau^-)$ (left) and $\sigma(pp\to\mu^-\mu^-\tau^+\tau^+)$ (right) for the LHC running at $\sqrt{s}=14$ TeV, and for different $m_{Z^\prime}$ benchmarks (cf. Table~\ref{tab:colliderbenchmarks}).
    The vertical lines denote the spread of the cross-section as allowed from scanning over the allowed parameter space accounting for $\Delta a_\mu$ (the interval of $g^{\mu\tau}_{LR}$ allowed for $m_{Z^\prime}$, cf. Table~\ref{tab:couplings_masses}). On the left, the coloured horizontal line denotes the SM prediction (and the associated MC uncertainties -- in paler shades).}
    \label{fig:LHC14_osl_ssl}
\end{figure}
In Fig.~\ref{fig:LHC14_osl_ssl} we display the results for the LHC cross-sections\footnote{Notice that $\sigma(pp\to\mu^+\mu^+\tau^-\tau^-)=\sigma(pp\to\mu^-\mu^-\tau^+\tau^+)$; hence, we only display the $\mu^-\mu^-\tau^+\tau^+$ case.} $\sigma(pp\to\mu^+\mu^-\tau^+\tau^-)$ and $\sigma(pp\to\mu^-\mu^-\tau^+\tau^+)$ for the considered $Z^\prime$ benchmarks (detailed in Table~\ref{tab:colliderbenchmarks}). 
The vertical (blue) lines denote the spread of the cross-section as predicted through scanning over the allowed parameter space accounting for $\Delta a_\mu$ (that is, the interval of $g^{\mu\tau}_{L,R}$ allowed for every choice of $m_{Z^\prime}$, cf. Table~\ref{tab:couplings_masses}). 

Notice that for the process 
$pp\to\mu^+\mu^-\tau^+\tau^-$, there are also SM contributions: in this case, the coloured horizontal line denotes the associated SM prediction (and the associated Monte Carlo (MC) uncertainties as quoted by {\tt MadGraph5\_aMC@NLO} -- in paler shades), whereas the full BSM result -- i.e. the vertical blue lines -- includes both SM and NP contributions.
Other than the difference arising from possible SM contributions (or absence thereof), the behaviour of the cross-sections is very similar for both processes, which reflects the fact that they share a similar underlying topology (Feynman diagrams such as the one shown in Fig.~\ref{fig:ffcollisions})
and thus involve very similar kinematics. In particular, the cross-section increases up to the benchmark $m_{Z^\prime} = 50$~GeV and its subsequent reduction stems from the fact that either the intermediate $Z$ or the $Z'$ will no longer be close to their respective poles.

\begin{figure}[h!]
    \centering
    \includegraphics[width=0.49\textwidth]{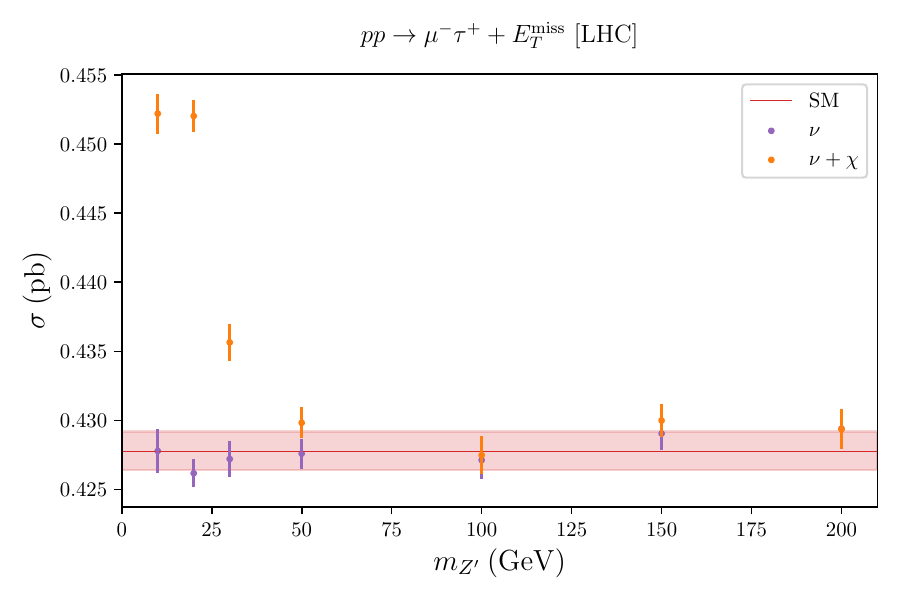}
    \caption{LHC: prospects for the cross-sections $\sigma(pp\to\mu^-\tau^+ + E^\text{miss}_T$), with $E^\text{miss}_T$ associated with  $\bar\nu_\mu \nu_\tau$, $\nu_\mu \bar\nu_\tau$ and $\chi\chi$, at the LHC running at $\sqrt{s}=14$ TeV (for different $m_{Z^\prime}$ benchmarks).
    The purple (orange) vertical lines denote $\mu^-\tau^+ \nu \nu$ ($\mu^-\tau^+ \nu \nu $ and $\mu^-\tau^+ \chi \chi$) final states. The spread of the lines corresponds to the uncertainty in the computation of the cross-sections (whose central values correspond to the coloured circles). As before the coloured horizontal line denotes the SM prediction (and the associated MC uncertainties -- in paler shades). }
    \label{fig:LHC14_inv}
\end{figure}

\medskip
In view of the role of the new gauge boson as a potential DM mediator, we now consider $\mu\tau$ pair production in association with missing energy. 
Signatures involving $E^\text{miss}_T$ can be due to three sources: the pure SM contribution (i.e. SM processes leading to a final state including  neutrinos, $\mu^-\tau^+ + E^\text{miss}_T$); $E^\text{miss}_T$ arising in the context of the present NP model but involving invisible final states only containing neutrinos; finally, due to the additional presence of dark matter in the final state, i.e. $\nu + \chi$.
Notice that for regimes in which one has a heavy dark matter candidate (i.e. $m_\chi > m_{Z^\prime}$), the contributions 
of the heavy $\chi$s in the final state are negligible, and hence the predictions are similar to the case in which $E^\text{miss}_T$ is solely composed of neutrinos.
In Fig.~\ref{fig:LHC14_inv} we display the cross-sections for
$ pp\to \mu^-\tau^+ + E^\text{miss}_T$, in which $E^\text{miss}_T$ is assumed to correspond to $\bar\nu_\mu \nu_\tau$, $\nu_\mu \bar\nu_\tau$ and $\chi\chi$.
We omit the corresponding results for heavy ($m_\chi > m_{Z^\prime}/2$) DM candidates, as they would lead  to the same predictions obtained by only considering neutrinos in the final state.

As manifest from Fig.~\ref{fig:LHC14_inv}, for the benchmarks  $Z^\prime_{10}\chi_{2}$, $Z^\prime_{20}\chi_{5}$ and $Z^\prime_{30}\chi_{10}$,
one is led to predictions for the associated cross-sections which are very distinct from those expected in the SM and from the NP ones with only neutrinos as missing energy. On the other hand, for final states in which the missing energy is solely due to the exotic (i.e. different-flavour) production of neutrinos, as well as for benchmarks featuring a heavier $Z^\prime$, the cross-sections are phenomenologically indistinguishable from the SM ones (regardless of the $Z^\prime$ mass).

\begin{figure}[h!]
    \centering
    \includegraphics[width=0.49\textwidth]{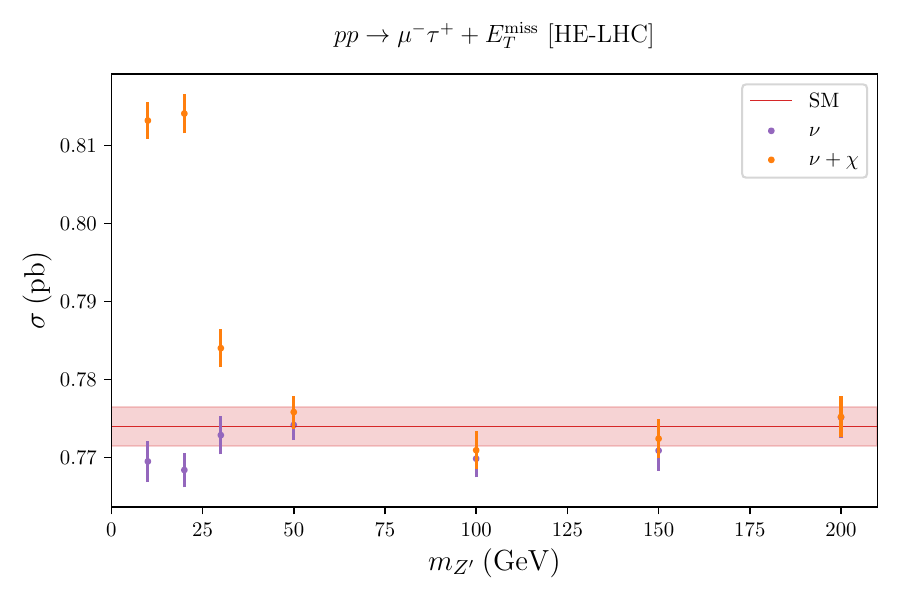} 
    \caption{HE-LHC: prospects for the cross-sections, $\sigma(pp\to\mu^-\tau^++ E_T^\mathrm{miss})$ for the HE-LHC running at $\sqrt{s}=27$ TeV, depicted for the considered $m_{Z^\prime}$ benchmark points.
    Colour code as in Fig.~\ref{fig:LHC14_inv}. }
    \label{fig:LHCHE_inv}
\end{figure}

\paragraph{HE-LHC}
The behaviour of the cross-sections, for both same- and opposite-sign dilepton pairs, is very similar to the one encountered for the LHC (running at $14$~TeV).
\begin{figure}[h!]
    \centering
    \includegraphics[width=0.49\textwidth]{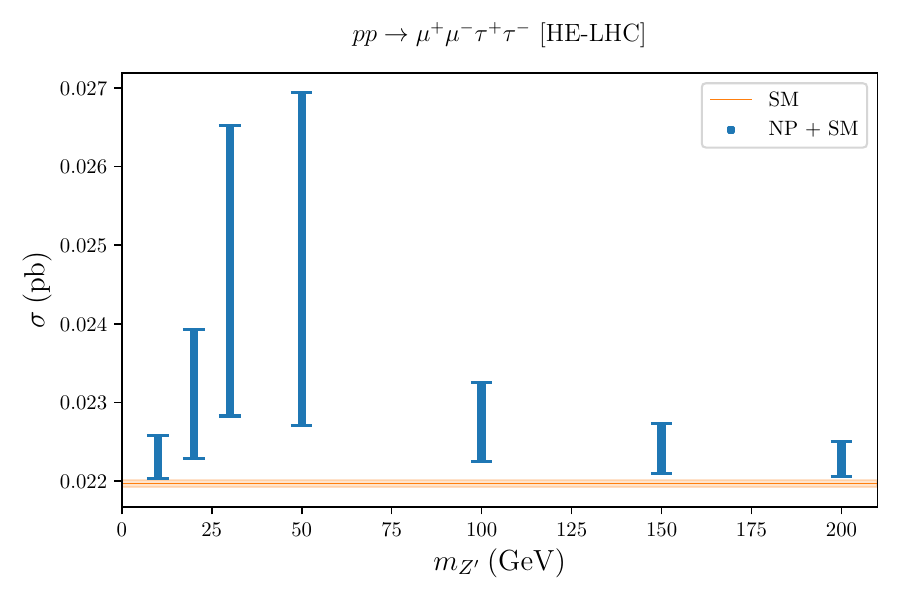} \hfill
    \includegraphics[width=0.49\textwidth]{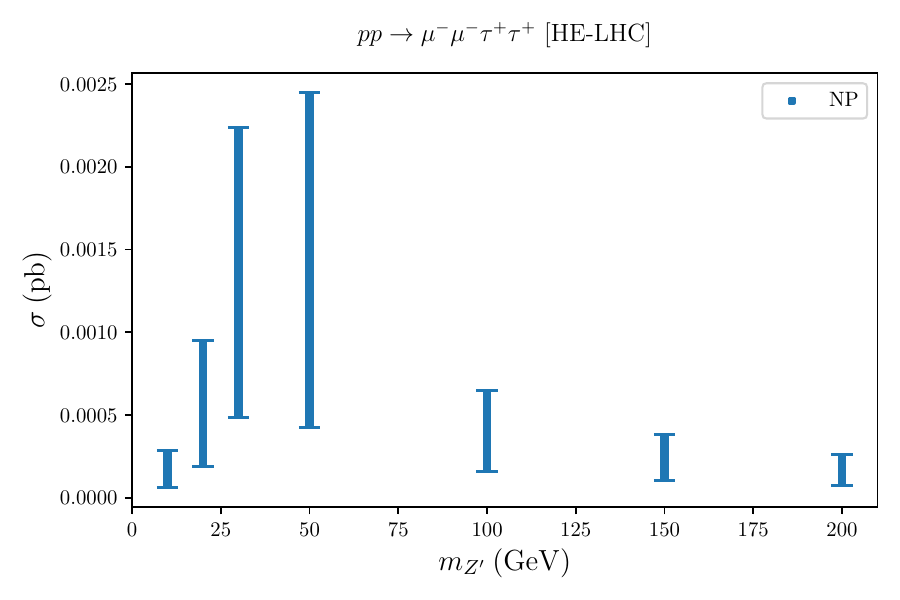} 
    \caption{HE-LHC: prospects for the cross-sections, $\sigma(pp\to\mu^+\mu^-\tau^+\tau^-)$ (left) and $\sigma(pp\to\mu^-\mu^-\tau^+\tau^+)$ (right) for the HE-LHC running at $\sqrt{s}=27$ TeV, depicted for the considered $m_{Z^\prime}$ benchmark points.
    Colour code as in Fig.~\ref{fig:LHC14_osl_ssl}. }
    \label{fig:LHCHE_osl_ssl}
\end{figure}
In Figs.~\ref{fig:LHCHE_osl_ssl} and~\ref{fig:LHCHE_inv} we display the cross-sections for the processes $pp\to\mu^+\mu^-\tau^+\tau^-$/$pp\to\mu^-\mu^-\tau^+\tau^+$ (Fig.~\ref{fig:LHCHE_osl_ssl}) and $pp\to\mu^-\tau^+ + E_T^\mathrm{miss}$ (Fig.~\ref{fig:LHCHE_inv}) for the HE-LHC run at $\sqrt{s}=27$ TeV. We observe that, similar to the case of the LHC, lower $Z^\prime$ mass scenarios could potentially be observed, especially in the case of same-sign dilepton pairs, with a similar remark holding for the low-mass range of the $\mu^-\tau^+ + E_T^\mathrm{miss}$ channel.

The higher cross-section values obtained in the case of HE-LHC (and this will be also manifest for the case of FCC-hh) 
can be attributed to the increasing  quark-antiquark content of protons for \textit{lower} values of $x$ (the parton momentum fraction): for a given kinematically allowed process (i.e. for given -- kinematically accessible -- values of the heavier particle masses involved in the process), higher $\sqrt{s}$ values correspond to lower $x$ values, and thus to a larger probability for $q\bar{q}$-initiated processes to occur.

\begin{figure}[h!]
    \centering
    \includegraphics[width=0.49\textwidth]{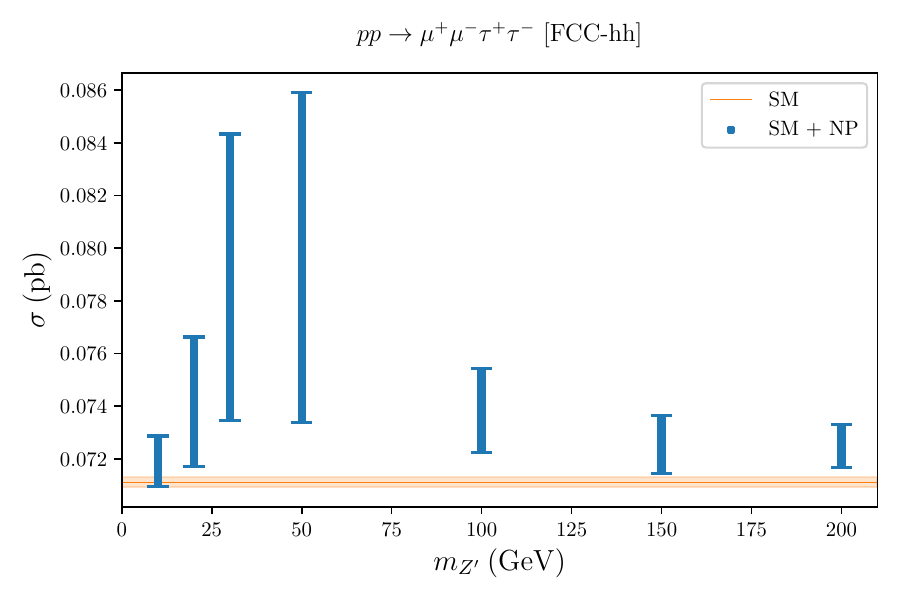} \hfill
    \includegraphics[width=0.49\textwidth]{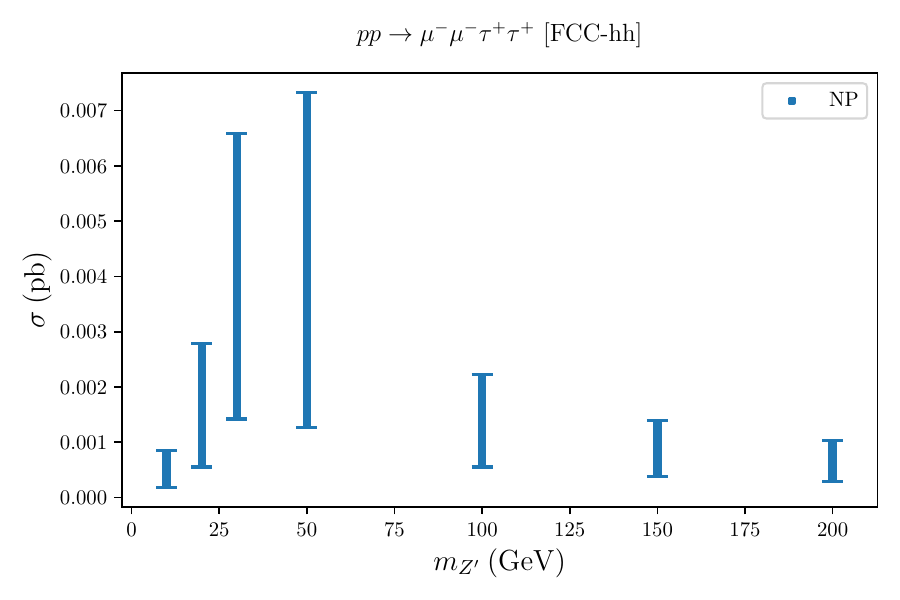} 
    \caption{FCC-hh: prospects for the cross-section $\sigma(pp\to\mu^-\mu^+\tau^-\tau^+)$ (left) and $\sigma(pp\to\mu^-\mu^-\tau^+\tau^+)$ (right) at a future FCC-hh running at $\sqrt{s}=100$ TeV, for the considered $m_{Z^\prime}$ benchmark points.
    Colour code as in Fig.~\ref{fig:LHC14_osl_ssl}. }
    \label{fig:FCChh_osl_ssl}
\end{figure}

\paragraph{FCC-hh} 
To conclude the analysis of hadron colliders, we now address the prospects of a future FCC-hh operating at a centre-of-mass energy of $\sqrt{s}=100$~TeV. 
In Fig.~\ref{fig:FCChh_osl_ssl} we first display the cross-sections $\sigma(pp\to\mu^+\mu^-\tau^+\tau^-)$ (left) $\sigma(pp\to\mu^-\mu^-\tau^+\tau^+)$ (right) for the different benchmark points.  
As expected, the behaviours are identical to the ones found for the LHC, with substantially larger associated cross-sections (by as much as a factor $~3$), both for opposite-sign and same-sign dilepton pairs.

\begin{figure}[h!]
    \centering
    \includegraphics[width=0.49\textwidth]{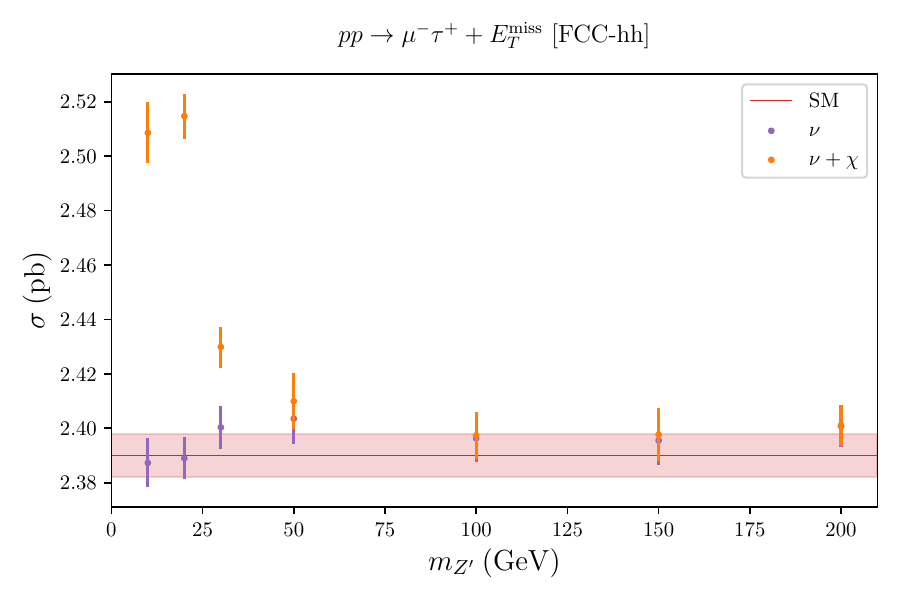} 
    \caption{FCC-hh: prospects for the cross-section $\sigma(pp\to\mu^-\tau^+ E_T^{\mathrm{miss}})$ for a future FCC-hh running at $\sqrt{s}=100$ TeV.
    Colour code as in Fig.~\ref{fig:LHC14_inv}. }
    \label{fig:FCChh_inv}
\end{figure}

For completeness, we also consider the associated missing energy signatures, $pp\to\mu^-\tau^+ + E_T^{\mathrm{miss}}$ for the FCC-hh, as shown in Fig.~\ref{fig:FCChh_inv}. With the exception of the lightest benchmark points (i.e. $m_{Z^\prime} \leq 50$~GeV and for missing energy associated to both neutrinos and dark matter), the NP contributions are hardly distinguishable from the SM expectation. 

\medskip

As already pointed out, the aim of this paper is not to perform a full assessment of the detectability of the considered processes for each collider setting. However, it is instructive to compare, for example, at least the minimal and maximal number of excess events that could be achieved in each type of facility with the expected background. Realistic background
estimates are far more complicated, involving 
among other factors, charge misidentification etc.
Here we do not take such effects into account.

In order to estimate the minimal and maximal number of BSM events that could be expected, we consider the minimum and maximum cross-section values that can be achieved for all of the benchmark scenarios presented in Table~\ref{tab:colliderbenchmarks} (i.e. across all considered $Z^\prime$ masses and couplings), focusing on the case $m_\chi < m_{Z^\prime}/2$, and convert 
these extremal values into expected events via the planned luminosities, as given in Table~\ref{tab:facility_energies}. 
In this manner we obtain $N_{\rm min/max}^{i}(j)$, where $i$ corresponds to the $i$-th collider configuration and $j$ to a given process. The uncertainty in the number of the SM background events will be denoted as $B^{i}(j)$, and corresponds to the na\"ive statistical uncertainty assuming Poisson statistics (equal to the square root of the number of events).

We find that 
only the most optimistic scenarios (i.e. those with the  highest cross-sections) could potentially be probed in hadron colliders. The most promising case is, as expected, the FCC-hh in which $N_{\rm min}^{\rm FCC-hh}(pp\to\mu^-\mu^+\tau^-\tau^+)$ is below the relevant SM background statistical uncertainty, whereas $N_{\rm max}^{\rm FCC-hh}(pp\to\mu^-\mu^+\tau^-\tau^+) \sim 300\times B^{\rm FCC-hh}(pp\to\mu^-\mu^+\tau^-\tau^+)$. 
A similar situation holds in the case of $pp\to\mu^-\tau^+ E_T^{\mathrm{miss}}$, for which the minimum value of the cross-section essentially leads to zero excess events, whereas $N_{\rm max}^{\rm FCC-hh}(pp\to\mu^-\tau^+ E_T^{\mathrm{miss}})$ can exceed the statistical uncertainty by up to a factor of $\sim 350$, corresponding to more than two million excess events. Same-sign processes typically lead to much smaller cross-sections but are essentially background-free and hence, should offer good detection possibilities throughout the benchmark regimes considered.
We note that these numbers have been obtained assuming an FCC-hh integrated luminosity of $20$ ab$^{-1}$.

\subsection{Future lepton colliders}
In view of the leptophilic nature of the new $Z^\prime$ mediator, future lepton colliders offer good prospects to study the production of opposite- and same-sign dilepton pairs. In this subsection, we explore all the envisaged possibilities: FCC-ee at different centre-of-mass energies, CLIC, a future high-energy muon collider and finally $\mu$TRISTAN.
As we will subsequently discuss, due to the underlying strictly $\mu-\tau$ flavour-violating nature of the $Z^\prime$ couplings, there are 
substantial similarities between the results for electron-positron colliders and the findings for hadron colliders.
Barring effects associated with the scaling of the quark-antiquark content of protons as a function of $x$ (the parton momentum fraction), for a given final state the parton-level cross-sections in $pp$ collisions are described by topologies similar to those relevant for $e^+ e^-$ collisions. 
On the other hand, muon colliders offer genuinely novel processes, and thus different opportunities to probe such NP models, as the underlying topologies are fundamentally different.

\begin{figure}[h!] 
    \centering
    \includegraphics[width=0.89\textwidth]{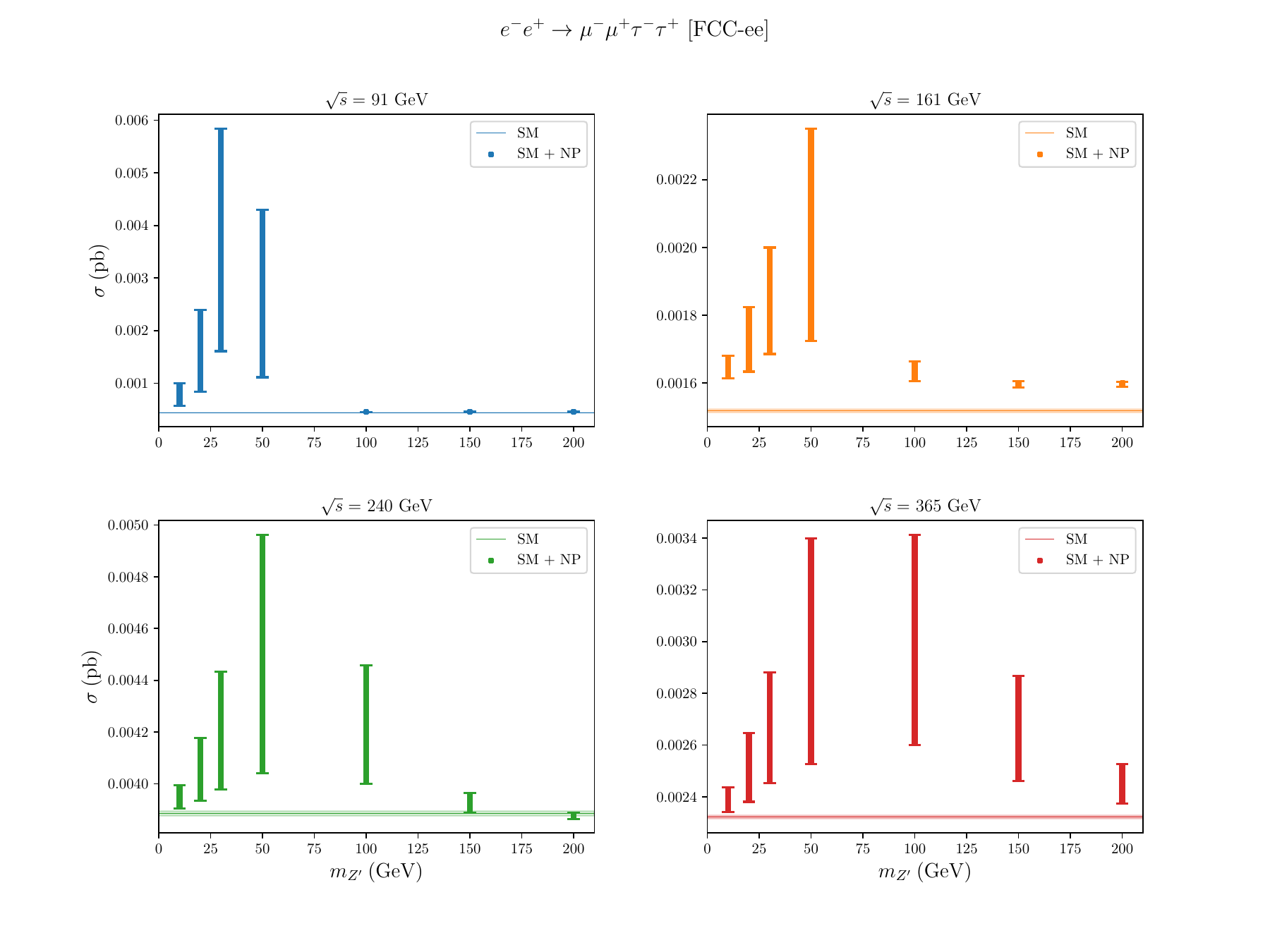} 
    \caption{FCC-ee: cross-section for $e^-e^+\to \mu^+\mu^-\tau^+\tau^-$ at FCC-ee for the considered $m_{Z^\prime}$
    benchmarks, and for distinct $\sqrt{s}$:
    $\sqrt{s}=91,\,161,\,240,\,365$ GeV (from left to right, top to bottom, respectively blue, orange, green and red).
    The vertical bands correspond to the spread of the cross-section allowed while accounting for $\Delta a_\mu$ (cf. Table~\ref{tab:couplings_masses}), while the horizontal lines denote the SM predictions.}
    \label{fig:FCCee_osl}
\end{figure}

\begin{figure}[h!] 
    \centering
    \includegraphics[width=0.89\textwidth]{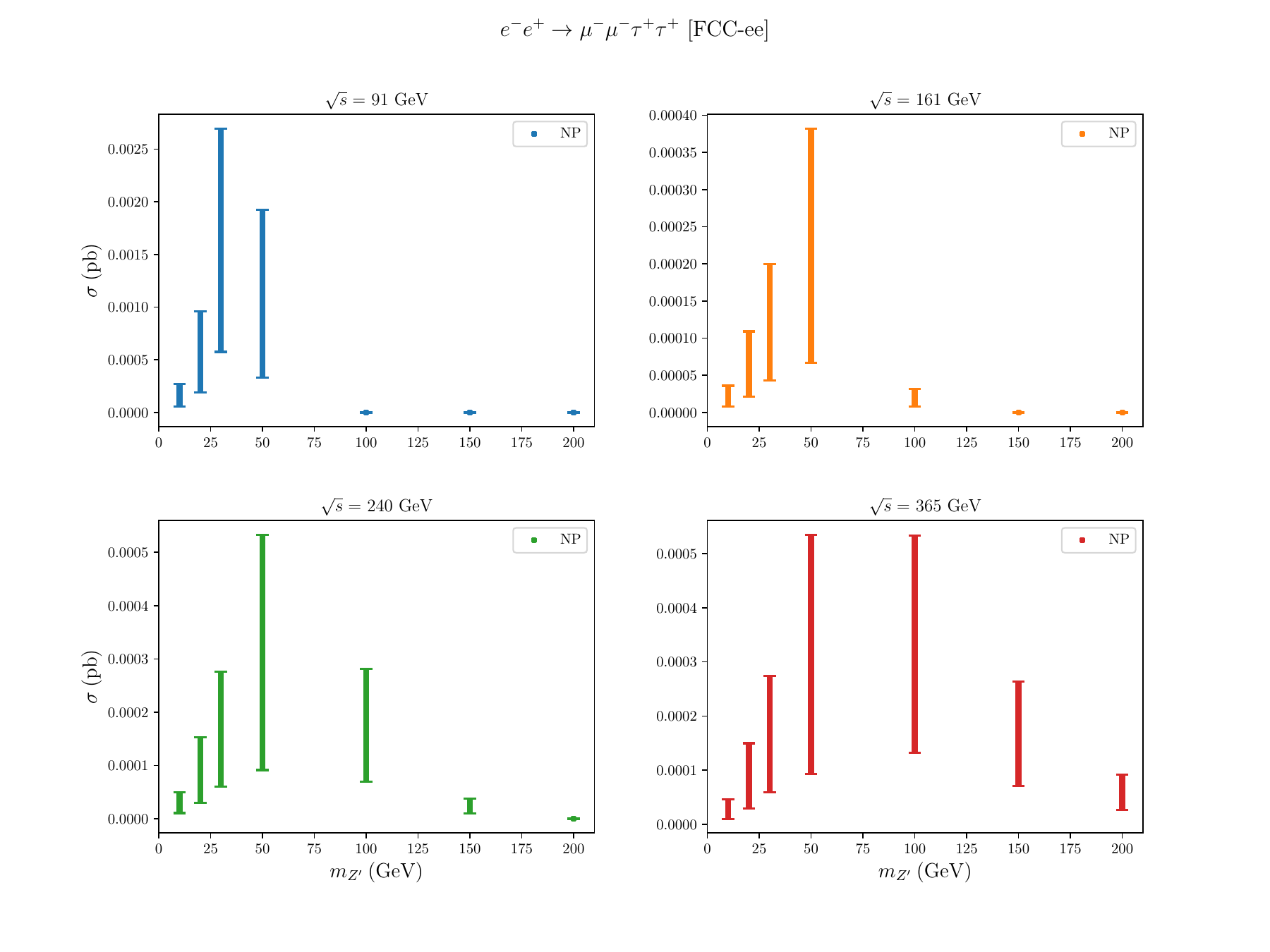}
    \caption{FCC-ee: cross-section for $e^-e^+\to \mu^-\mu^-\tau^+\tau^+$ at FCC-ee for the considered $m_{Z^\prime}$
    benchmarks and different $\sqrt{s}$. Line and colour code as in Fig.~\ref{fig:FCCee_osl}.}
    \label{fig:FCCee_ssl}
\end{figure}

\paragraph{FCC-ee}
We start the analysis of future lepton colliders with FCC-ee. We display the cross-sections for the processes $e^-e^+\to \mu^+\mu^-\tau^+\tau^-$ and $e^-e^+\to \mu^-\mu^-\tau^+\tau^+$ in Figs.~\ref{fig:FCCee_osl} and~\ref{fig:FCCee_ssl}, respectively. In both figures, we present the cross-sections for the $m_{Z^\prime}$ considered benchmarks, and for the different FCC-ee planned centre-of-mass energy runs: $\sqrt{s}=91,\,161,\,240,\,365$ GeV (from left to right, top to bottom in both figures). We recall that an example of a possible topology for the processes under scrutiny was given in Fig.~\ref{fig:ffcollisions}.

For each value of the centre-of-mass energy, and as long as $Z^\prime$ production is kinematically accessible, the cross-sections for opposite- and same-sign dilepton pairs exhibit a similar behaviour (similar to the findings for hadron colliders, discussed in the previous subsection), as expected from the same underlying topology of the contributing diagrams. Across different centre-of-mass energies, we first note that for $\sqrt{s} = 91$ GeV the cross-sections can be substantially enhanced due to having an on-shell $Z$ boson mediating the process in the $s$-channel. Moreover, the cross-section for $\sqrt{s}=240$ GeV is further boosted in the case of opposite-sign dilepton pairs, since both the $Z$ and the $Z^\prime$ can be on-shell. All in all, the magnitude of both the SM background and that of the BSM contribution varies depending on the interplay between different contributions to the total cross-section. Moreover, for $\sqrt{s}=91$ GeV, the cross-section $\sigma(e^-e^+\to \mu^+\mu^-\tau^+\tau^-)$ is manifestly distinguishable from the SM expectations for $m_{Z^\prime}\leq 50$ GeV. At larger centre-of-mass energies, the FCC-ee would have the potential to probe most of the regimes associated with the benchmark points proposed in Table~\ref{tab:colliderbenchmarks}.

\begin{figure}[h!] 
    \centering
    \includegraphics[width=0.99\textwidth]{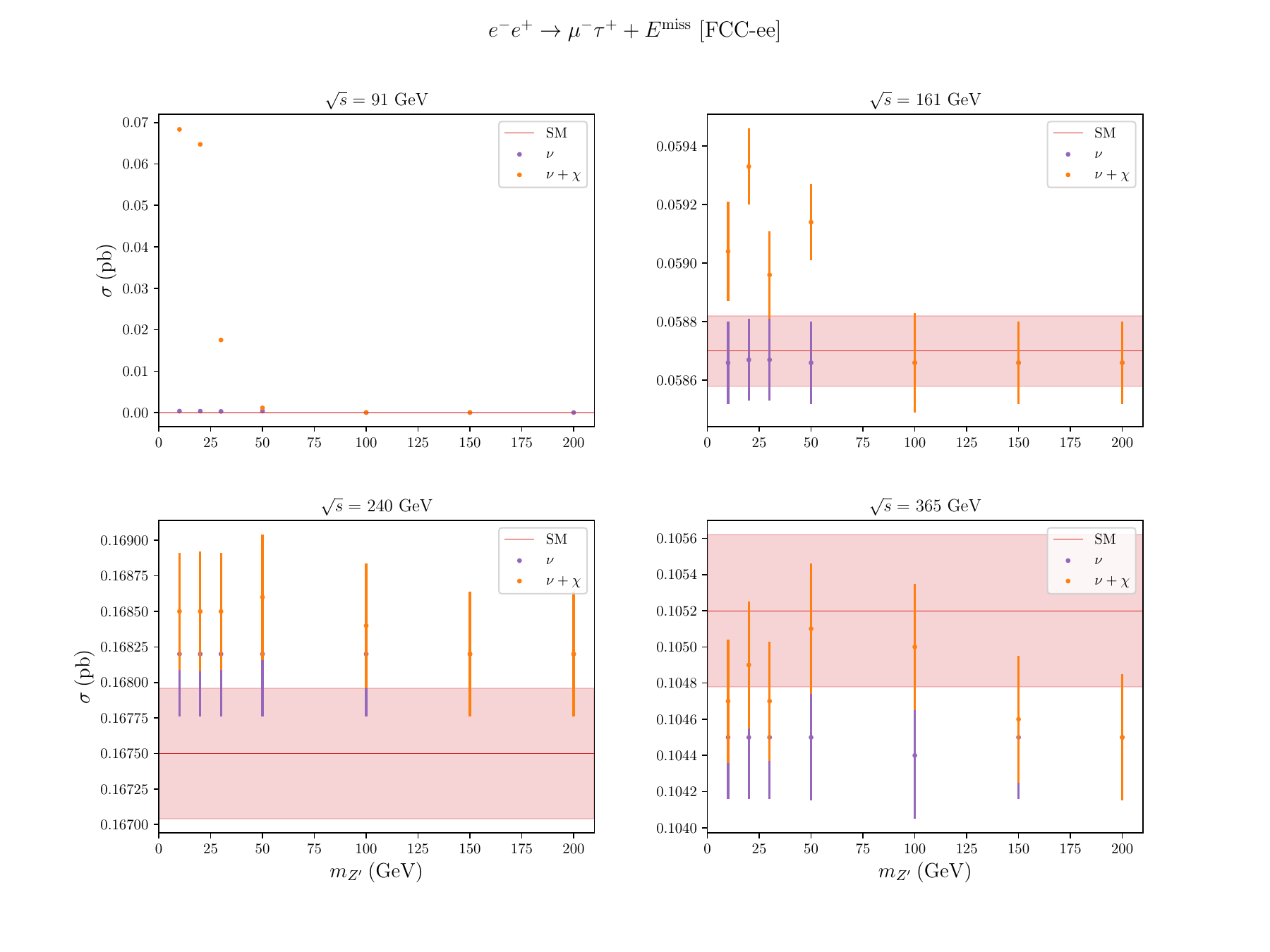} 
    \caption{FCC-ee: cross-section for $e^-e^+\to \mu^-\tau^+ + E_\text{miss}$  at FCC-ee  for the considered $m_{Z^\prime}$
    benchmarks and for different $\sqrt{s}$. Line and colour code as in Fig.~\ref{fig:LHC14_inv}.}
    \label{fig:FCCee_inv}
\end{figure}

For  signatures with missing energy, we display in Fig.~\ref{fig:FCCee_inv} the expected cross-sections for $e^-e^+\to \mu^-\tau^+ + E_\text{miss}$, in which $E_\text{miss}$  can be associated with neutrinos or correspond to $\nu + \chi$ emission.
As already pointed out in connection with the prospects for hadron colliders,  at $\sqrt{s}=91$~GeV the (DM) benchmark points $Z^\prime_{10}\chi_{2}$, $Z^\prime_{20}\chi_{5}$ and $Z^\prime_{30}\chi_{10}$ lead to cross-sections very distinct from the SM expectation, and from the case in which neutrinos account for all of the missing energy. 
This clearly confirms how the presence of light DM could induce sizeable deviations in the cross-section, potentially identifiable at the FCC-ee.
Although not as sizeable, we recover the same increase at $\sqrt{s}=161$~GeV, while at higher centre-of-mass energy the contributions from light DM tend to be negligible compared to the neutrino ones. Interestingly, and for $\sqrt{s}=365$~GeV, the (destructive) interference between the NP contributions and the SM ones leads to cross-sections which are considerably \textit{lower} than for the SM case (no new physics contributions). As seen from the lower right hand-side panel of Fig.~\ref{fig:FCCee_inv}, 
one could in principle be sensitive to such a minimal NP model from the searches for $e^-e^+\to \mu^-\tau^+ + E_\text{miss}$.

\begin{figure}[h!] 
    \centering
    \includegraphics[width=\textwidth]{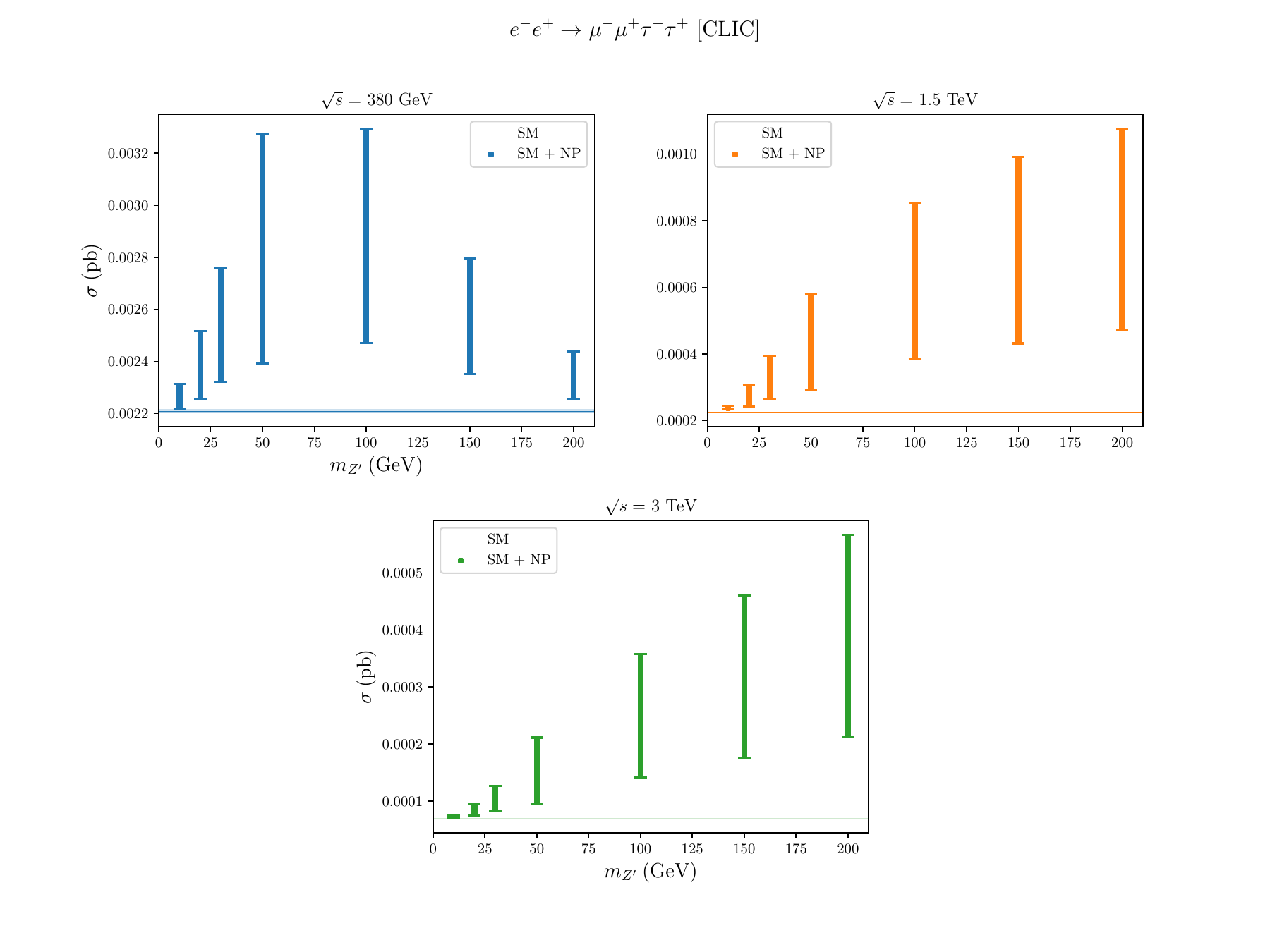}
    \caption{CLIC: cross-sections for $e^-e^+\to \mu^+\mu^-\tau^+\tau^-$ for the considered $m_{Z^\prime}$
    benchmarks at CLIC (for different $\sqrt{s}$).  Line and colour code as in Fig.~\ref{fig:FCCee_osl}.}
    \label{fig:CLIC_osl}
\end{figure}

\begin{figure}[h!] 
    \centering
    \includegraphics[width=0.49\textwidth]{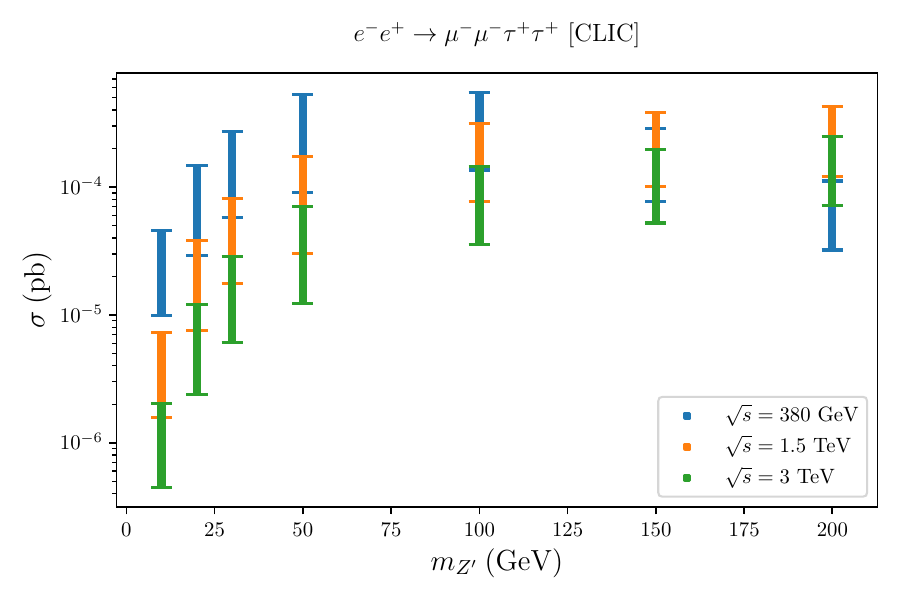}
    \caption{CLIC: cross-section for  $e^-e^+\to \mu^-\mu^-\tau^+\tau^+$ at CLIC for the considered $m_{Z^\prime}$
    benchmarks and for different $\sqrt{s}$.  Line and colour code as in Fig.~\ref{fig:FCCee_osl}.}
    \label{fig:CLIC_ssl}
\end{figure}

\paragraph{CLIC}
Linear colliders also offer excellent prospects to study lepton flavour violating NP models. In our study we thus investigate the prospects for CLIC: the cross-sections for $e^-e^+\to \mu^+\mu^-\tau^+\tau^-$ and $e^-e^+\to \mu^-\mu^-\tau^+\tau^+$ are shown in Figs.~\ref{fig:CLIC_osl} and~\ref{fig:CLIC_ssl}
(for different centre-of-mass energies).

As visible from Fig.~\ref{fig:CLIC_osl}, the behaviour of the cross-section for $e^-e^+\to \mu^+\mu^-\tau^+\tau^-$ does indeed follow the same evolution for comparatively low values of the centre-of-mass energy\footnote{This is a consequence of the cuts imposed on the momenta of the final state leptons; should these be relaxed or absent, one would see an increase of the cross-section with $m_{Z^\prime}$, similar to what is observed for $\sqrt s =1.5$~TeV and 3~TeV.} (see the panel corresponding to 
$\sqrt s =350$~GeV); however, it becomes very distinctive 
for higher collider energies (i.e. at 1.5~TeV and 3~TeV), for which it keeps increasing with the $Z^\prime$ mass (and couplings). Moreover for TeV centre-of-mass energies, the cross-section can be up to 5 times larger than the one predicted in the SM, thus suggesting an important discovery potential.

\begin{figure}[h!] 
    \centering
    \includegraphics[width=0.99\textwidth]{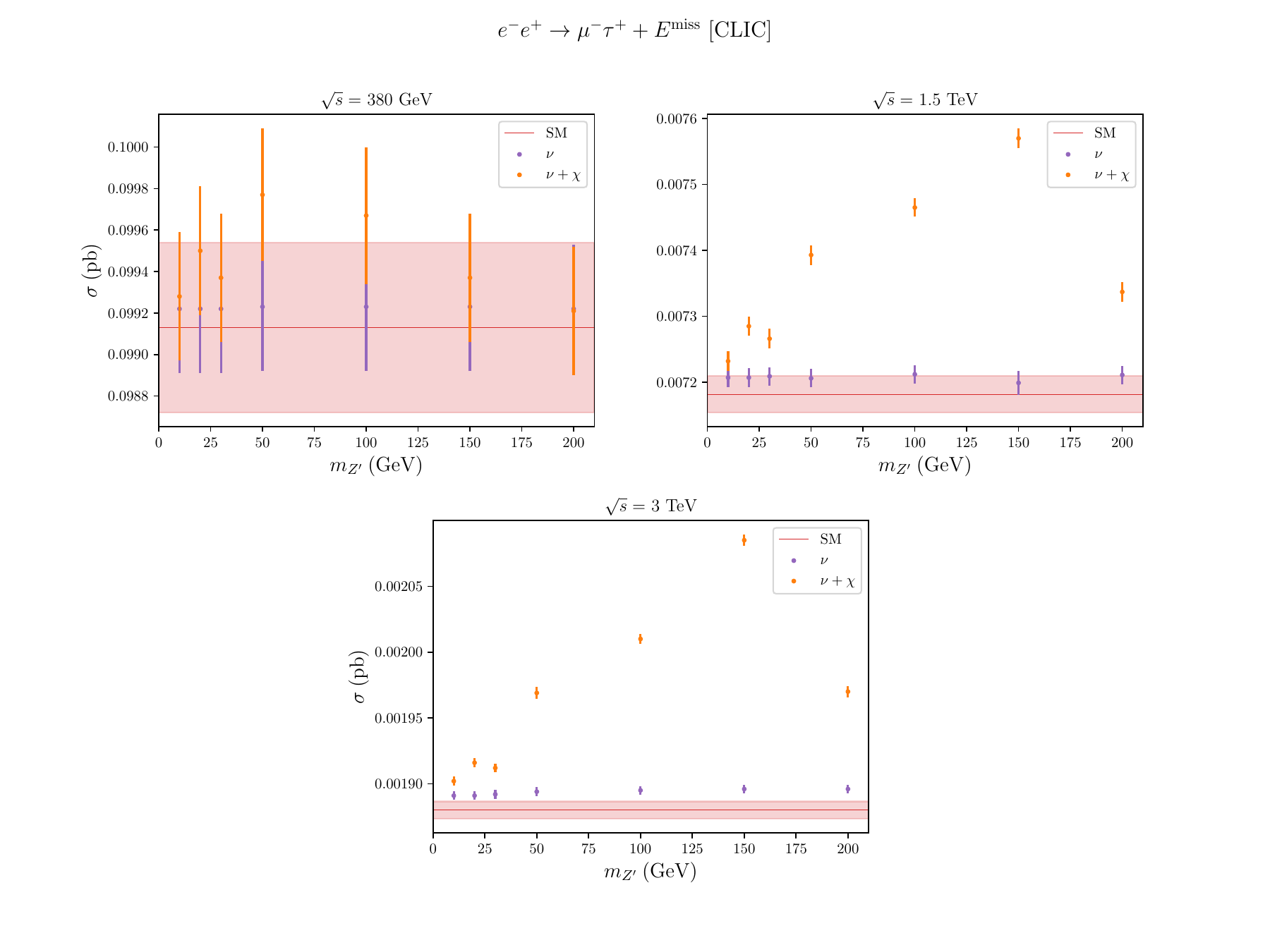}
    \caption{CLIC: cross-section for $e^-e^+\to \mu^-\tau^+ + E_\text{miss}$ at CLIC for the considered $m_{Z^\prime}$
    benchmarks and different $\sqrt{s}$. Line and colour code as in Fig.~\ref{fig:LHC14_inv}.}
    \label{fig:CLIC_inv}
\end{figure}

For final states consisting of same-sign dilepton pairs (see Fig.~\ref{fig:CLIC_ssl}), not only are the values of 
$\sigma(e^-e^+\to \mu^-\mu^-\tau^+\tau^+)$ considerably smaller -- by as much as two orders of magnitude -- but one also finds a unique behaviour of the cross-section for varying $m_{Z^\prime}$:
the contributions significantly increase with increasing new boson mass.
Carrying out the same strategy as before, let us now consider final states involving missing energy. This is displayed in Fig.~\ref{fig:CLIC_inv}, in which one can verify that for centre-of-mass energies above 1~TeV the cross-sections for final states involving light DM are larger than the missing energy signatures due to the presence of only neutrinos. Hence for the considered benchmarks, one could possibly identify the presence of light DM.  

\medskip
To conclude this survey of future electron colliders, we briefly comment on the ``na\"ive'' expected number of events (similar to what was done in the previous subsection). 
The best prospects at $e^+e^- $ colliders are found for the FCC-ee running at the $Z$-pole, both for opposite-sign dilepton pairs and when missing energy is involved in the final state. Indeed, one finds  $N_{\rm min}^{\text{ FCC-ee\,91}}(e^+e^-\to\mu^-\mu^+\tau^-\tau^+) > 4 B^{\text{FCC-ee\,91}}(e^+e^-\to\mu^-\mu^+\tau^-\tau^+)$, with a maximal number of excess events which can reach more than 3500 times the SM uncertainty (i.e. more than one million excess events). Moreover in the case of  $e^-e^+\to \mu^-\tau^+ + E_\text{miss}$ at 91 GeV FCC-ee, one could expect more than 13 million excess events considering the maximal cross-section. Even if $N_{\rm min}^{\text{ FCC-ee\,91}}(e^+e^-\to\mu^-\tau^+ + E_\text{miss})$ is roughly twice the relevant SM background statistical uncertainty, most of the considered parameter space (as represented via the chosen benchmark points) could be probed at the FCC-ee running at the $Z$-pole.

\begin{figure}[h!]
    \centering
\begin{tikzpicture}
    \begin{feynman}
    \vertex (mu1) at (0,1.2) {\(\mu^-\)};
    \vertex (mu2) at (0,-1.2){\( \mu^+\)};
    \vertex (a) at (2,1.2) ;
    \vertex (b) at (2,-1.2);
    \vertex (c) at (5,1.2) {\(\mu^-\)};
    \vertex (d) at (5,-1.2){\(\tau^+\)}; 
    \vertex (e) at (3,1.2);
    \vertex (f) at (3.7,0);
    \vertex (f1) at (5,0.4){\small\(\mu^- \)};
    \vertex (f2) at (5,-0.4){\small\(\tau^+ \)};
    \diagram* {
    (mu1) -- [fermion] (a) -- [fermion, edge label=\(\tau^-\)] (e) -- [fermion] (c),
    (a) -- [boson, edge label'=\(Z^\prime\)] (b),
    (mu2) -- [anti fermion] (b) -- [anti fermion] (d),
    (e) -- [boson, edge label'=\(Z^\prime\)] (f),
    (f1) -- [fermion](f) -- [fermion](f2)};
    \end{feynman}
\end{tikzpicture}
\hspace*{10mm}
\begin{tikzpicture}
    \begin{feynman}
    \vertex (mu1) at (0,1.2) {\(\mu^-\)};
    \vertex (mu2) at (0,-0.8){\( \mu^+\)};
    \vertex (a) at (1.7,1.2) ;
    \vertex (b) at (1.7,-0.8);
    \vertex (c) at (3,1.2) ;
    \vertex (d) at (3,-0.8); 
    \vertex (f1) at (4.5,0.8){\small\(\mu^- \)};
    \vertex (f2) at (4.5,1.6){\small\(\tau^+ \)};
    \vertex (f1a) at (4.5,-0.4){\small\(\mu^- \)};
    \vertex (f2a) at (4.5,-1.2){\small\(\tau^+ \)};
    \diagram* {
    (mu1) -- [fermion] (a) -- [fermion, edge label'=\(\tau^-\)] (b) -- [fermion] (mu2),
    (a) -- [boson, edge label'=\(Z^\prime\)] (c),
    (b) -- [boson, edge label'=\(Z^\prime\)] (d),
    (f1a) -- [fermion](d) -- [fermion](f2a),
    (f1) -- [fermion](c) -- [fermion](f2)};
    \end{feynman}
\end{tikzpicture}
    \caption{Illustrative example of a possible topology for the considered processes at a future $\mu^+\mu^-$ collider.}
    \label{fig:mumucollisions}
\end{figure}
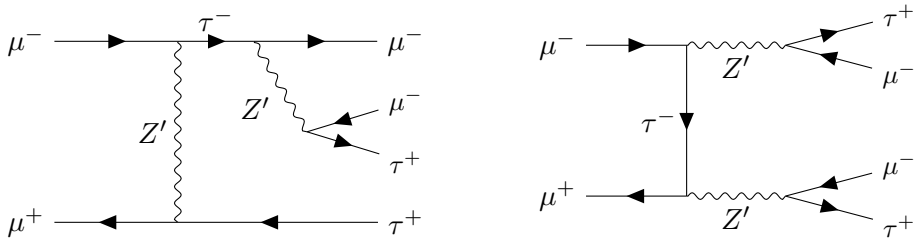

\paragraph {Muon colliders}
A future muon collider might offer the perfect environment to test (and possibly discover) models of NP with sizeable cLFV $\mu-\tau$ interactions, as is the present case. The processes are distinct to what was considered in previous paragraphs, as we consider $\mu \mu \to \mu \mu \tau \tau$ production. 
Figure~\ref{fig:mumucollisions} highlights such differences (additional diagrams, to better illustrate the processes at a muon collider, are presented in Appendix~\ref{app:feynman}). 

\begin{figure}[h!]
    \centering
    \includegraphics[width=0.49\textwidth]{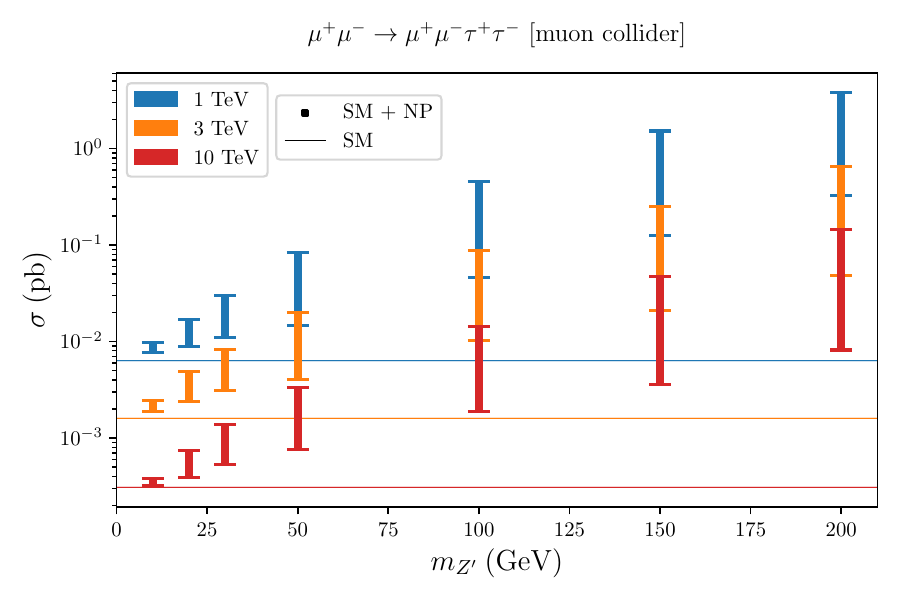}
    \hfill
    \includegraphics[width=0.49\textwidth]{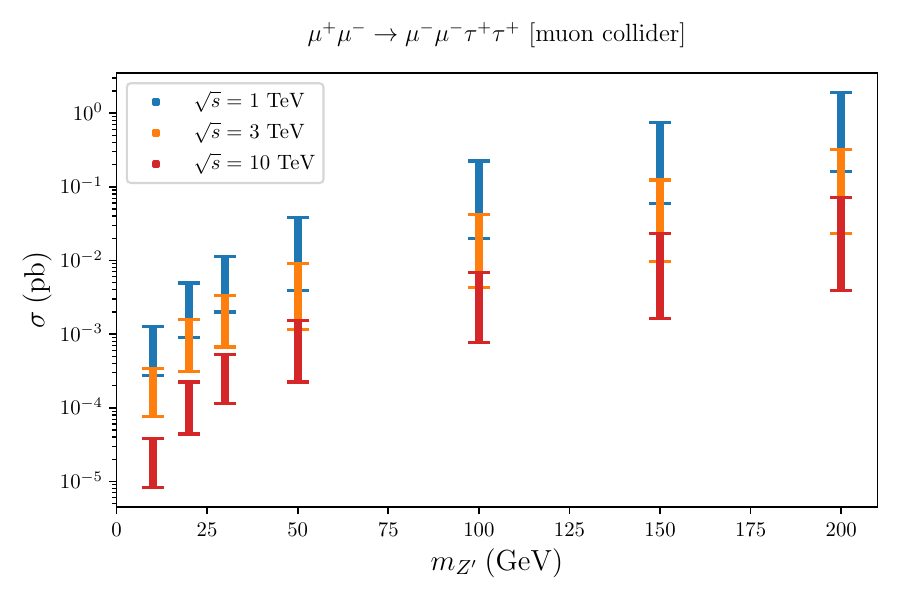}
    \caption{Muon collider: prospects for  $\sigma(\mu^+\mu^-\to \mu^- \mu^+ \tau^- \tau^+)$ (left) and $\sigma(\mu^+\mu^-\to \mu^- \mu^- \tau^+ \tau^+)$ (right)  for several $m_{Z^\prime}$ benchmarks at a future muon collider. The line code is as in Fig.~\ref{fig:LHC14_osl_ssl}, and the colour scheme denotes the considered centre-of-mass energies 1~TeV (blue), 3~TeV (orange) and 10~TeV (red).}
    \label{fig:mucoll:ssl_osl}
\end{figure}

In Fig.~\ref{fig:mucoll:ssl_osl} we display the cross-sections for same-sign final states, i.e. $\mu^+\mu^-\to \mu^- \mu^- \tau^+ \tau^+$, and opposite-sign final states $\mu^+\mu^-\to \mu^- \mu^+ \tau^- \tau^+$, for centre-of-mass energies of 1~TeV, 3~TeV and 10~TeV. As can be seen, the production cross-sections are very large (rendering very likely the observation of a significant number of events), especially for heavier values of $m_{Z^\prime}$. 
Although the cross-sections for same-sign dilepton pairs are
slightly smaller, one can also expect a large number of events, 
corresponding to a striking BSM signature.

\begin{figure}[h!]
    \centering
    \includegraphics[width=0.49\textwidth]{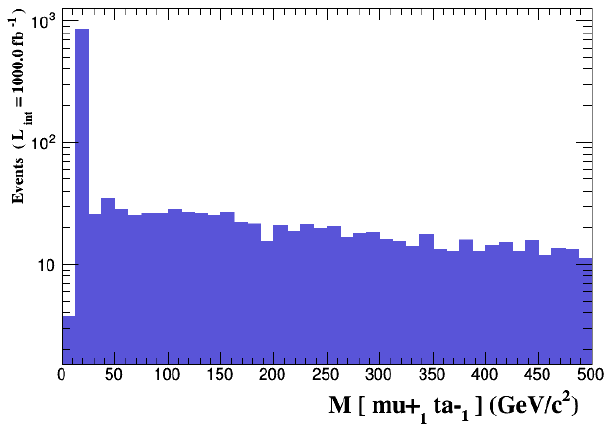}
    \hfill \includegraphics[width=0.49\textwidth]{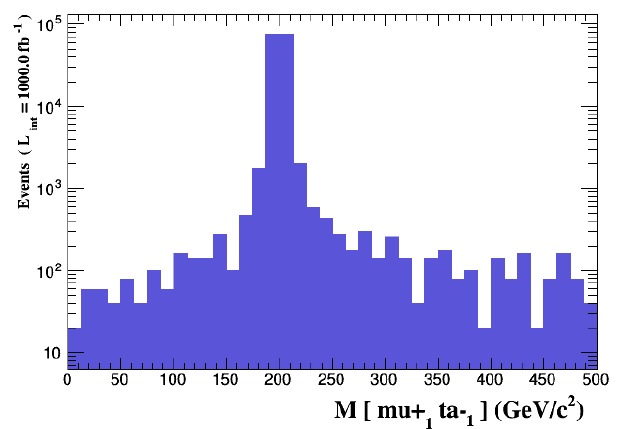}
    \caption{Muon collider: dilepton invariant mass distributions, $M(\mu^+\tau^-)$, for $m_{Z^\prime}= 20$~GeV (left) and 200~GeV (right) associated with $\mu^+\mu^-\to \mu^+ \mu^- \tau^+ \tau^-$ at a 3~TeV future muon collider.}
    \label{fig:mucoll:Mzp:peak:osl}
\end{figure}

\medskip
One can also investigate the dilepton invariant mass distribution,  $M(\mu^-\tau^+)$,  in SM allowed process such as $\mu^+\mu^-\to \mu^+ \mu^- \tau^+ \tau^-$, looking for peaks reflecting the presence of a $Z^\prime$ boson. Our findings for the dilepton invariant mass distributions are presented in Fig.~\ref{fig:mucoll:Mzp:peak:osl}, leading to which we have considered $m_{Z^\prime}=20$~GeV and $m_{Z^\prime}=200$~GeV (for a muon collider operating at $3$~TeV).
As manifest from both plots, the resonant production of the 
$Z^\prime$ leads to very clear signatures. 
Although we do not include the results here, we notice that the same clear peaks, corresponding to 
on-shell $Z^\prime$ production and decay, have been found for all considered benchmarks, and for centre-of-mass 
energies of 1~TeV and 10~TeV. 

\begin{figure}[h!]
    \centering
    \includegraphics[width=0.99\textwidth]{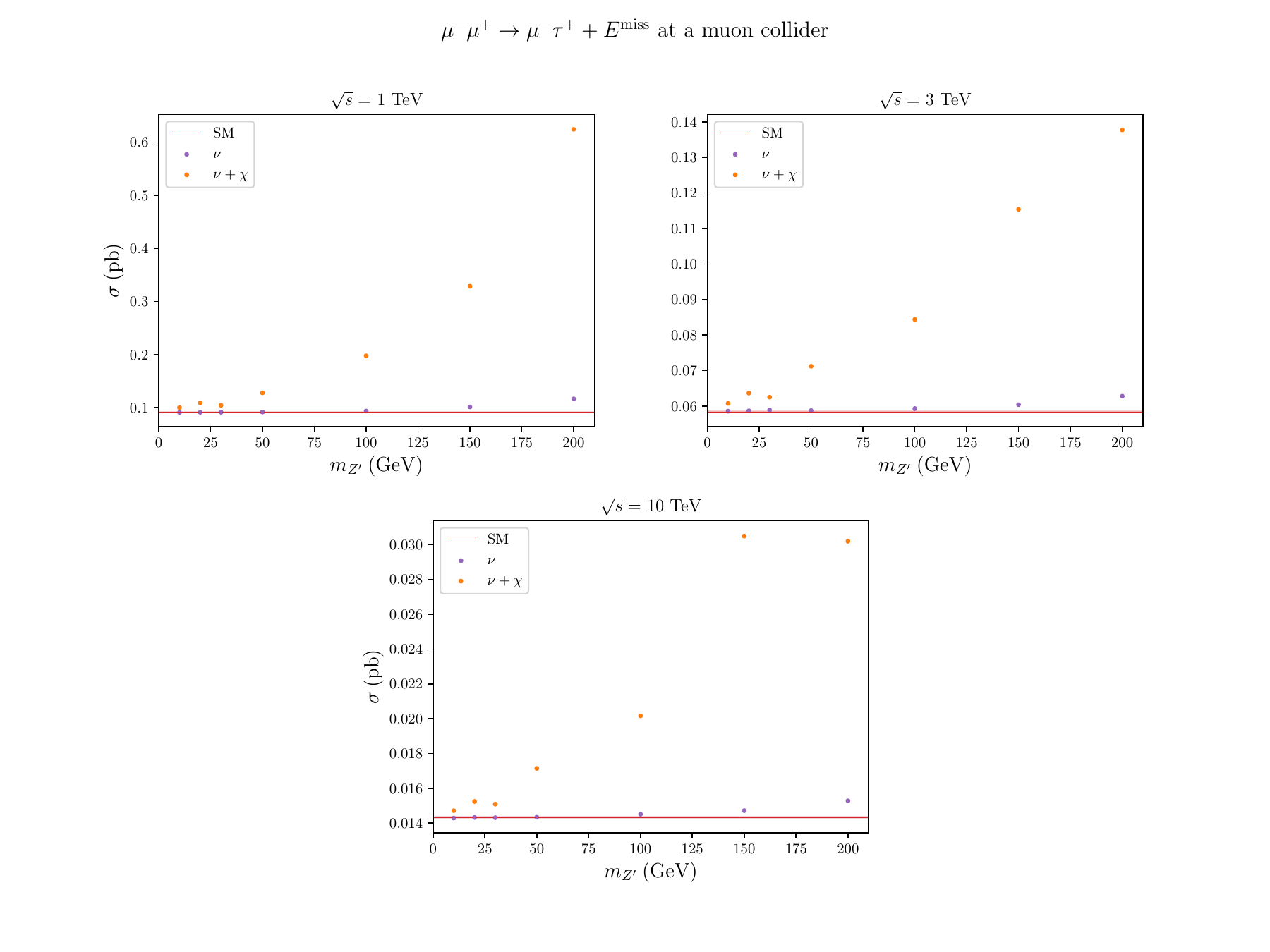}
    \caption{Muon collider: $\sigma(\mu^+\mu^-\to \tau^- \tau^+ + E^\text{miss}$) for the considered $m_{Z^\prime}$
    benchmarks and for different centre-of-mass energies at a future muon collider. Line and colour code as in Fig.~\ref{fig:LHC14_inv}.}
    \label{fig:mucoll:inv}
\end{figure}

\begin{figure}[h!]
    \centering
    \includegraphics[width=0.99\textwidth]{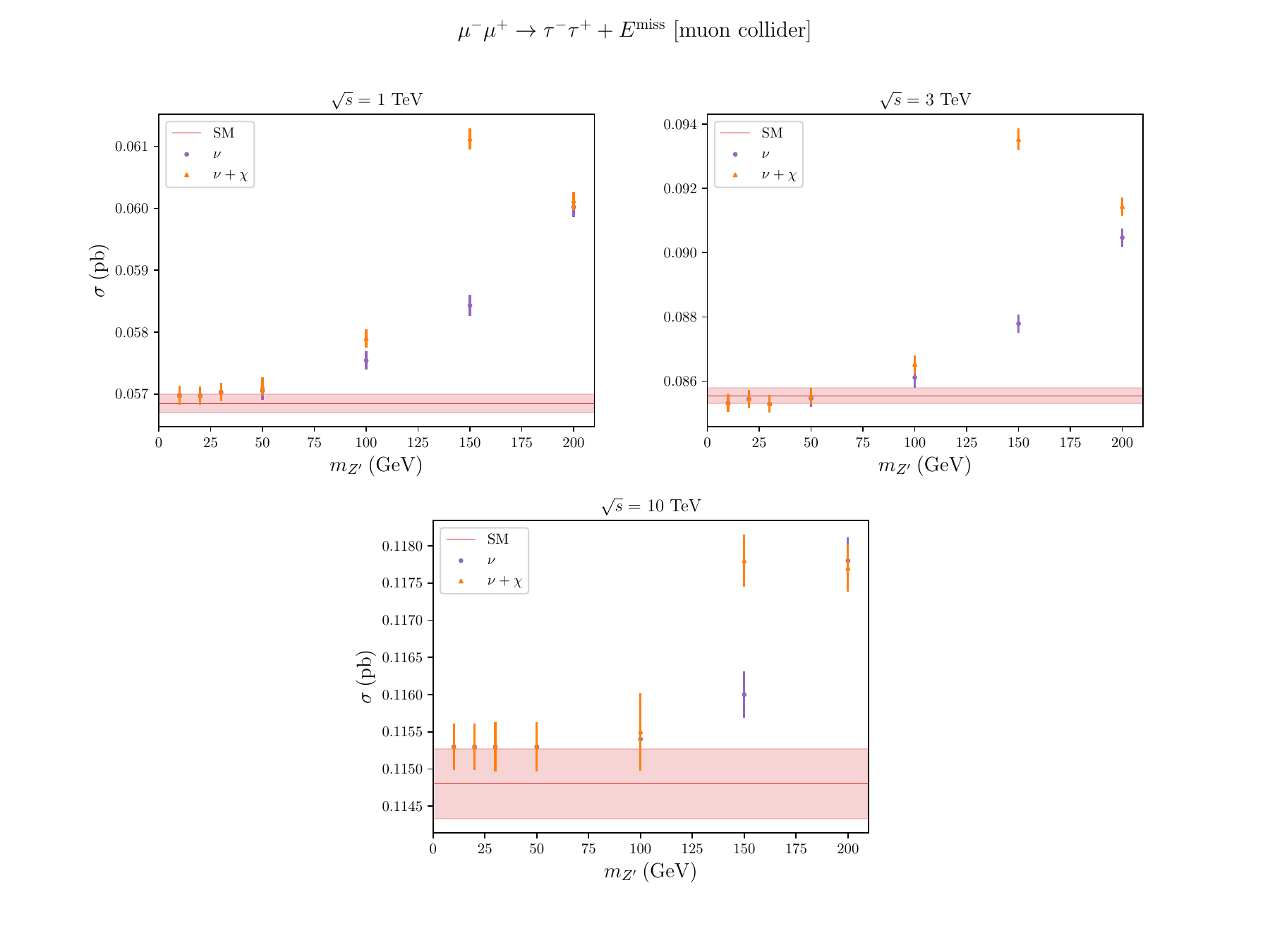}
    \caption{Muon collider: $\sigma(\mu^+\mu^-\to \tau^- \tau^+ + E^\mathrm{miss}$) for the considered $m_{Z^\prime}$
    benchmarks and for different centre-of-mass energies at a future muon collider. Line and colour code as in Fig.~\ref{fig:LHC14_inv}.}
    \label{fig:mucoll:tt:inv}
\end{figure}

We also consider final states with missing energy for the case of a muon collider, and our results are collected in Figs.~\ref{fig:mucoll:inv} and~\ref{fig:mucoll:tt:inv}, respectively for  $\mu^+\mu^-\to \mu^- \tau^+ + E^\text{miss}$ and $\mu^+\mu^-\to \tau^- \tau^+ + E^\text{miss}$. For the first case\footnote{Notice that in this case, and although we formally display vertical bars for the predictions associated with each benchmark, their spread is not visible to the naked eye.}, the prospects for NP signatures with missing energy due to neutrinos are not very promising (hard to distinguish from the SM background); however, for a hybrid 
$E^\text{miss}$ content (i.e. neutrinos and DM), the prospects are extremely good, both regarding the large cross-section as well as a clear distinction from the SM expectation, as occurs for benchmarks with heavier $Z^\prime$. For the second case (Fig.~\ref{fig:mucoll:tt:inv}), the prospects for observing a distinctive NP signature in association with missing energy are promising -- irrespective of the nature of the underlying  particles escaping detection. A particular interesting configuration emerges for a muon collider operating at $\sqrt s=3$~TeV, with sizeable cross-sections and a clear signal-background separation.

\paragraph{$\mu$TRISTAN}
To conclude the prospective study of the impact of such a minimal NP construction for future colliders, we consider $\mu$TRISTAN, which offers the possibility of $\mu^+\mu^+$ and  $e^-\mu^+$ colliding beams,  leading to very diversified scenarios regarding the possible final state. Peculiar topologies to this future facility are collected in Appendix~\ref{app:feynman}. 

\begin{figure}[h!]
    \centering
    \includegraphics[width=0.49\textwidth]{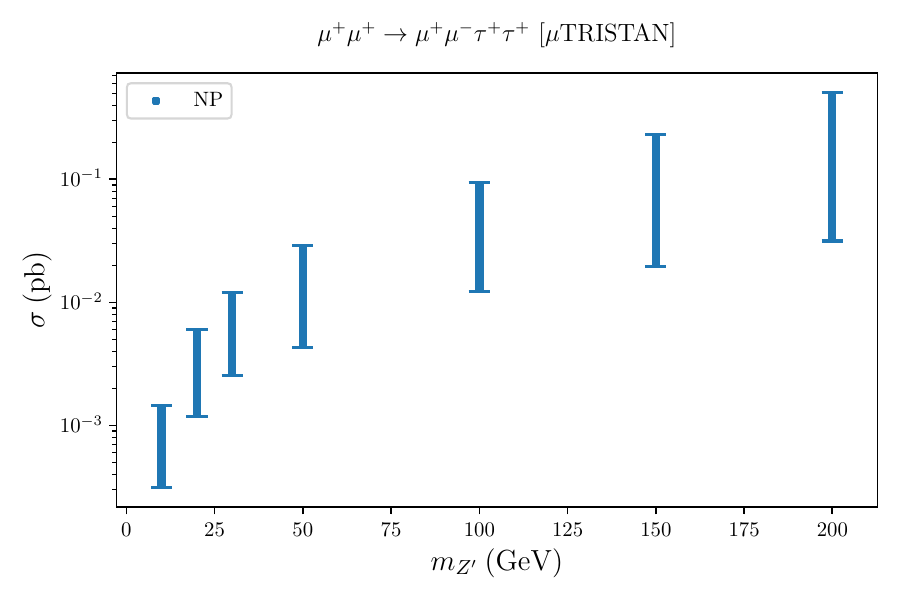}\hfill
    \includegraphics[width=0.49\textwidth]{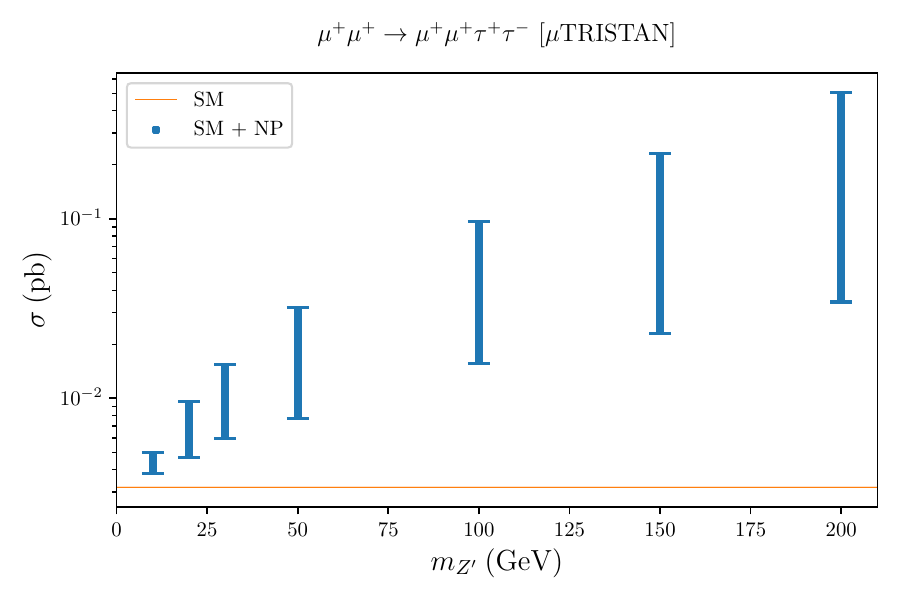}
    \caption{$\mu$TRISTAN: cross-section for the process $\mu^+\mu^+\to \mu^+ \mu^- \tau^+ \tau^+$ (left) and $\mu^+\mu^+\to \mu^+ \mu^+ \tau^- \tau^+$ (right) for the considered $m_{Z^\prime}$
    benchmarks at $\mu$TRISTAN. Line and colour code as in Fig.~\ref{fig:FCCee_osl}.}
    \label{fig:Tristan_2_osmu_ostau}
\end{figure}
Beginning with $\mu^+\mu^+$ beams, in Fig.~\ref{fig:Tristan_2_osmu_ostau} we thus display the different cross-sections for the NP contributions to the process $\mu^+\mu^+\to \mu^+ \mu^- \tau^+ \tau^+$ and $\mu^+\mu^+\to \mu^+ \mu^+ \tau^- \tau^+$ at $\mu$TRISTAN (running at $\sqrt{s}=2$ TeV).  Possibly, the cross-sections for benchmarks associated with larger values of $m_{Z^\prime}$ may render viable the observation of such process in the future.

\begin{figure}[h!]
    \centering
    \includegraphics[width=0.49\textwidth]{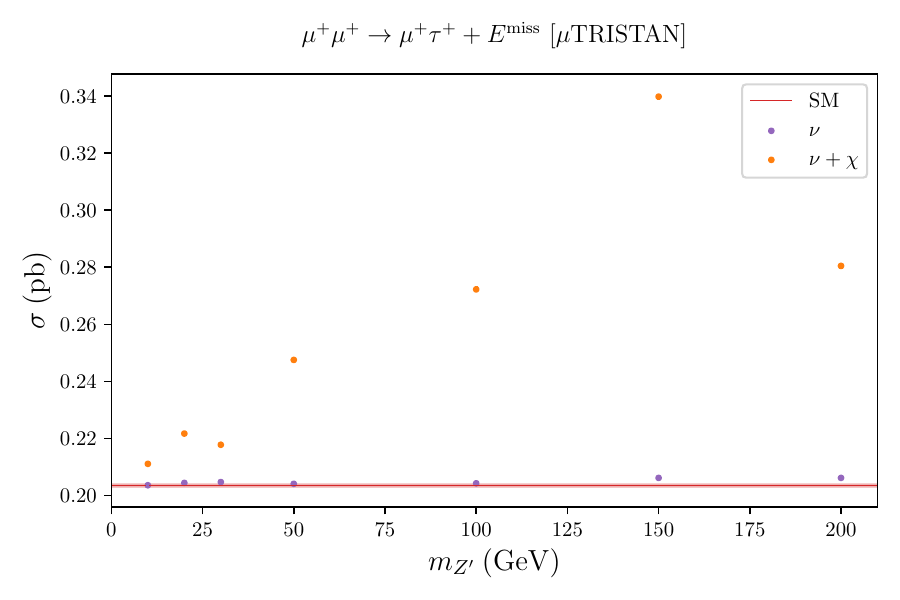}
      \includegraphics[width=0.49\textwidth]{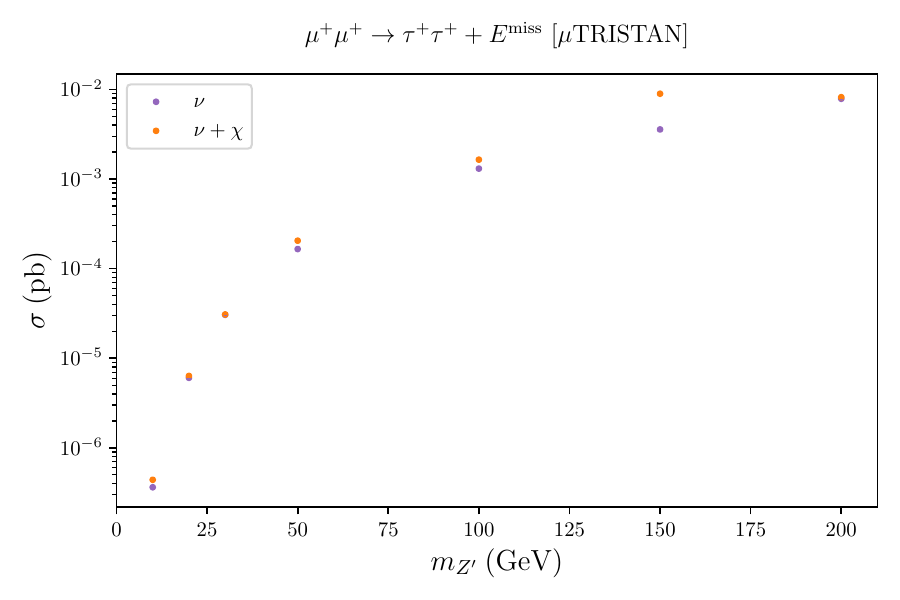}
        \caption{$\mu$TRISTAN ($\mu^+\mu^+$): cross-section for the process $\mu^+\mu^+\to \mu^+ \tau^+ + E^\text{miss}$ (left) and for the process $\mu^+\mu^+\to \tau^+ \tau^+ + E^\text{miss}$ (right)
    for the considered $m_{Z^\prime}$
    benchmarks at $\mu$TRISTAN. Line and colour code as in Fig.~\ref{fig:LHC14_inv}.}
    \label{fig:Tristan_2_inv}
\end{figure}

\begin{figure}[h!]
    \centering
    \includegraphics[width=0.49\textwidth]{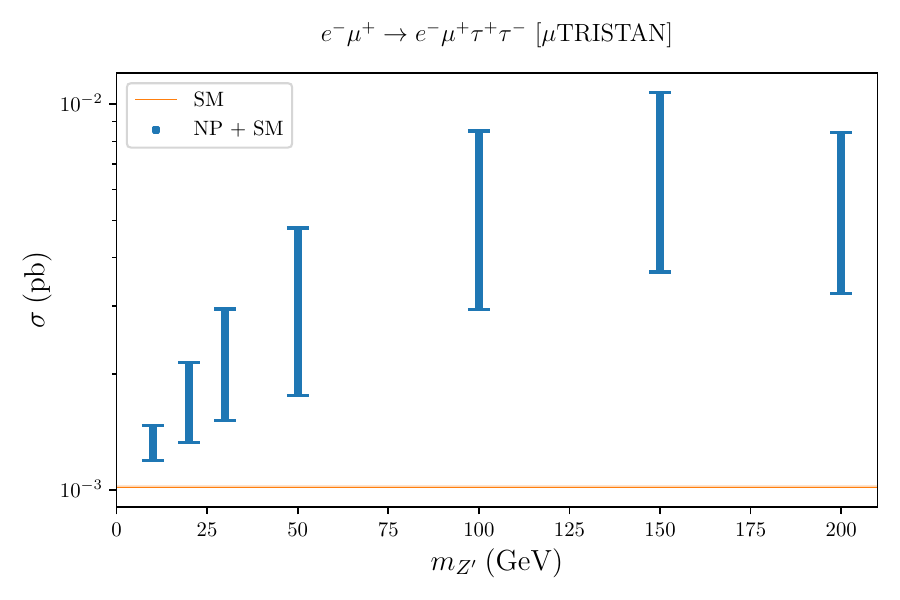} \hfill
    \includegraphics[width=0.49\textwidth]{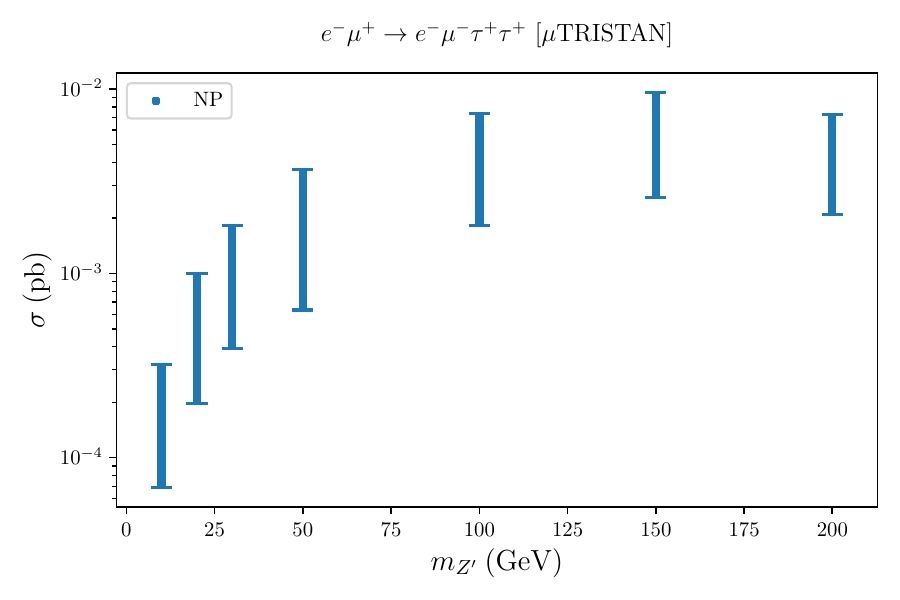}
    \caption{$\mu$TRISTAN ($e^-\mu^+$): $\sigma(e^-\mu^+\to e^-\mu^+\tau^+\tau^-)$ (left) and $\sigma(e^-\mu^+\to e^-\mu^-\tau^+\tau^+)$ (right) for the considered $m_{Z^\prime}$
    benchmarks at $\mu$TRISTAN. Line and colour code as in Fig.~\ref{fig:FCCee_osl}.}
    \label{fig:Tristan_emu346_ostau_sstau}
\end{figure}

Our study also envisages searches for processes leading to missing energy from $\mu^+\mu^+$ collisions: in Fig.~\ref{fig:Tristan_2_inv} 
we  respectively present the different cross-sections for the SM and NP contributions to the process $\mu^+\mu^+\to \mu^+ \tau^+ + E^\text{miss}$ and $\mu^+\mu^+\to \tau^+ \tau^+ +E^\text{miss}$, at $\mu$TRISTAN (running at $\sqrt{s}=2$ TeV).
Especially for the $\mu^+ \tau^+ + E^\text{miss}$ final states, and again for regimes associated with ``large'' $m_{Z^\prime}$, the sizeable cross-sections expected at $\mu$TRISTAN suggest that these decays will be observable.
For $\mu$TRISTAN operating in its $e^-\mu^+$ mode, the predictions for the SM and NP contributions to the cross-sections
$\sigma(e^-\mu^+\to e^-\mu^+\tau^+\tau^-)$ and $\sigma(e^-\mu^+\to e^-\mu^-\tau^+\tau^+)$ are summarised in Fig.~\ref{fig:Tristan_emu346_ostau_sstau}. Notice however that the associated values for the crosss section are more modest in this case.

\begin{figure}[h!]
    \centering
    \includegraphics[width=0.69\textwidth]{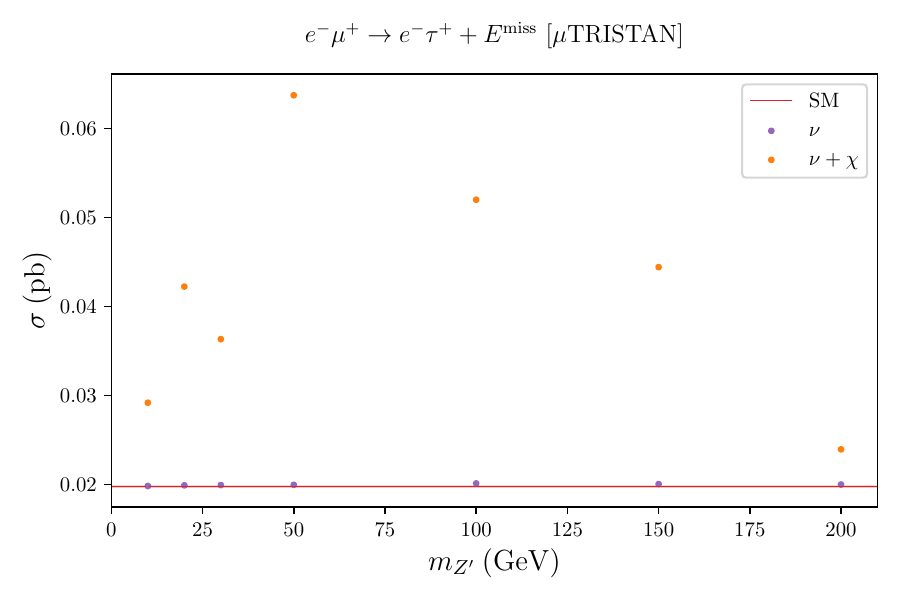}
    \caption{$\mu$TRISTAN ($e^-\mu^+$): $\sigma(e^-\mu^+\to e^-\mu^+\tau^+\tau^-)$ (left) and $\sigma(e^-\mu^+\to e^-\mu^-\tau^+\tau^+)$ (right) for the considered $m_{Z^\prime}$
    benchmarks at $\mu$TRISTAN. Line and colour code as in Fig.~\ref{fig:LHC14_inv}.}
    \label{fig:Tristan_emu346_inv}
\end{figure}

Likewise, Fig.~\ref{fig:Tristan_emu346_inv} displays the cross-section for the missing energy modes, $e^-\mu^+\to e^-\tau^+ + E^\text{miss}$ at $\mu$TRISTAN (which are smaller than the expectations for the previously discussed  $\mu^+\mu^+\to \mu^+ \tau^+ + E^\text{miss}$ in association with the $\mu^+\mu^+$ mode).

\medskip

Finally, and to emphasise the very good (if not excellent) prospects of a future muon collider, we discuss the number of excess events that could be expected. 
For 3~TeV muon collisions, we estimate $N_{\rm min}^{\mu \rm{Coll}\,3}(\mu^+\mu^-\to\mu^-\mu^+\tau^-\tau^+) > 7 B^{\mu \mathrm{Coll}\,3}(\mu^+\mu^-\to\mu^-\mu^+\tau^-\tau^+)$, while the maximum number of excess event is $\sim 640 000$. The signature $\mu^+\mu^-\to\mu^-\tau^+ + E^\mathrm{miss}$ would also lead to an important number of events, more than 9 times the statistical estimated background, and up to more than 79000 excess events in the case of the maximal cross-section of the explored regions. The prospects for $\mu^+\mu^-\to\tau^-\tau^+ + E^\mathrm{miss}$ are however not as good, as one can also have a deficit in the number of events (with respect to the SM expectation) which would lie within the SM background statistical uncertainty. Hence only benchmarks 
leading to larger cross-sections could be probed.

Even if less spectacular than what is observed for a future muon collider, it is worth noticing that the expected number of NP events at $\mu$TRISTAN, for both its $\mu^+\mu^+$ and $e^-\mu^+$ modes, is in all cases 
larger than the statistic fluctuations (for opposite-sign ditau and for missing energy in the final state).

\subsection{Appraisal for collider prospects}
In the previous subsections we have considered various processes (with final states consisting either of 4-leptons or of 2-leptons accompanied by missing energy) for a variety of present and future high-energy colliders. 
Comparing the prospects across distinct experimental setups is not a straightforward task, especially bearing in mind that some of the processes are fundamentally different, be it due to the nature of the colliding beams or to the composition of the final state. 

\begin{figure}[h!]
    \centering
    \includegraphics[width=0.49\textwidth]{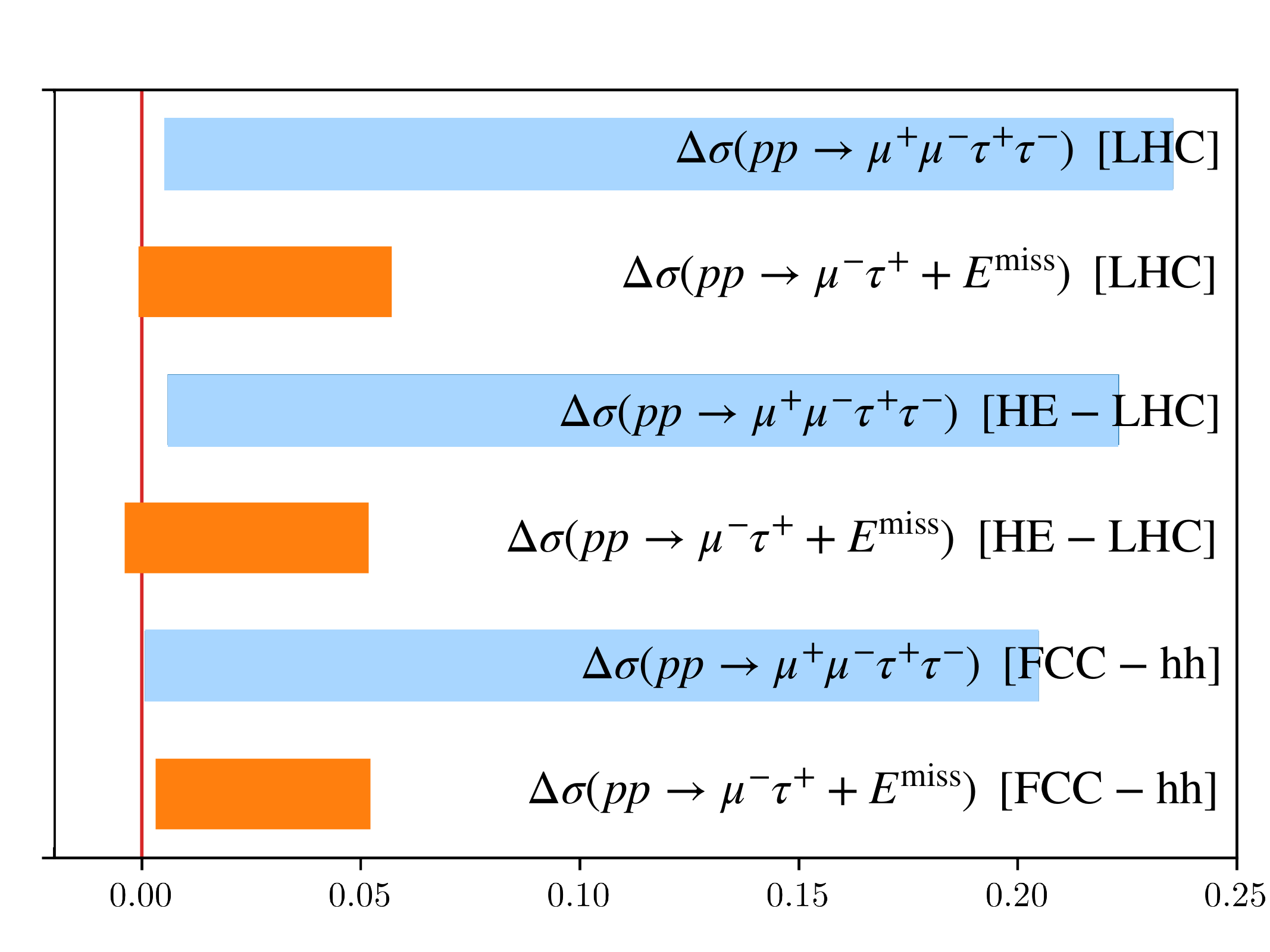}
    \caption{Comparison of the prospects for hadron colliders: $\Delta \sigma^{pp}_{Z^\prime \chi}$ (LHC and FCC-hh).}
    \label{fig:delta_hadron}
\end{figure}

\begin{figure}[h!]
    \centering
    \includegraphics[width=0.89\textwidth]{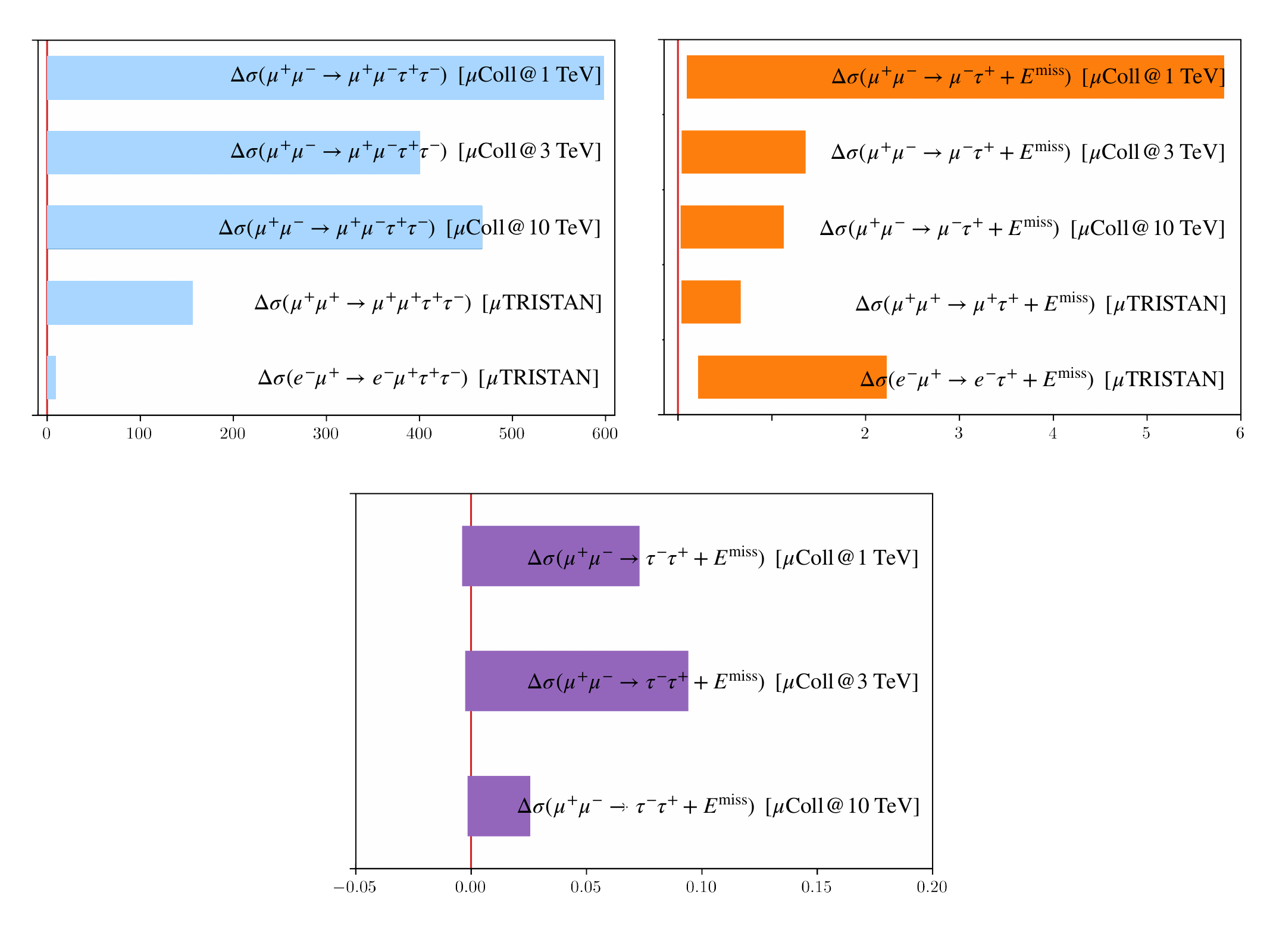}
    \caption{Comparison of the prospects for muon colliders: $\Delta \sigma^{\mu\mu}_{Z^\prime \chi}$ (future muon collider and $\mu$TRISTAN at the different $\sqrt s$ runs).}
    \label{fig:delta_mucoll}
\end{figure}

\begin{figure}[h!]
    \centering
    \includegraphics[width=0.89\textwidth]{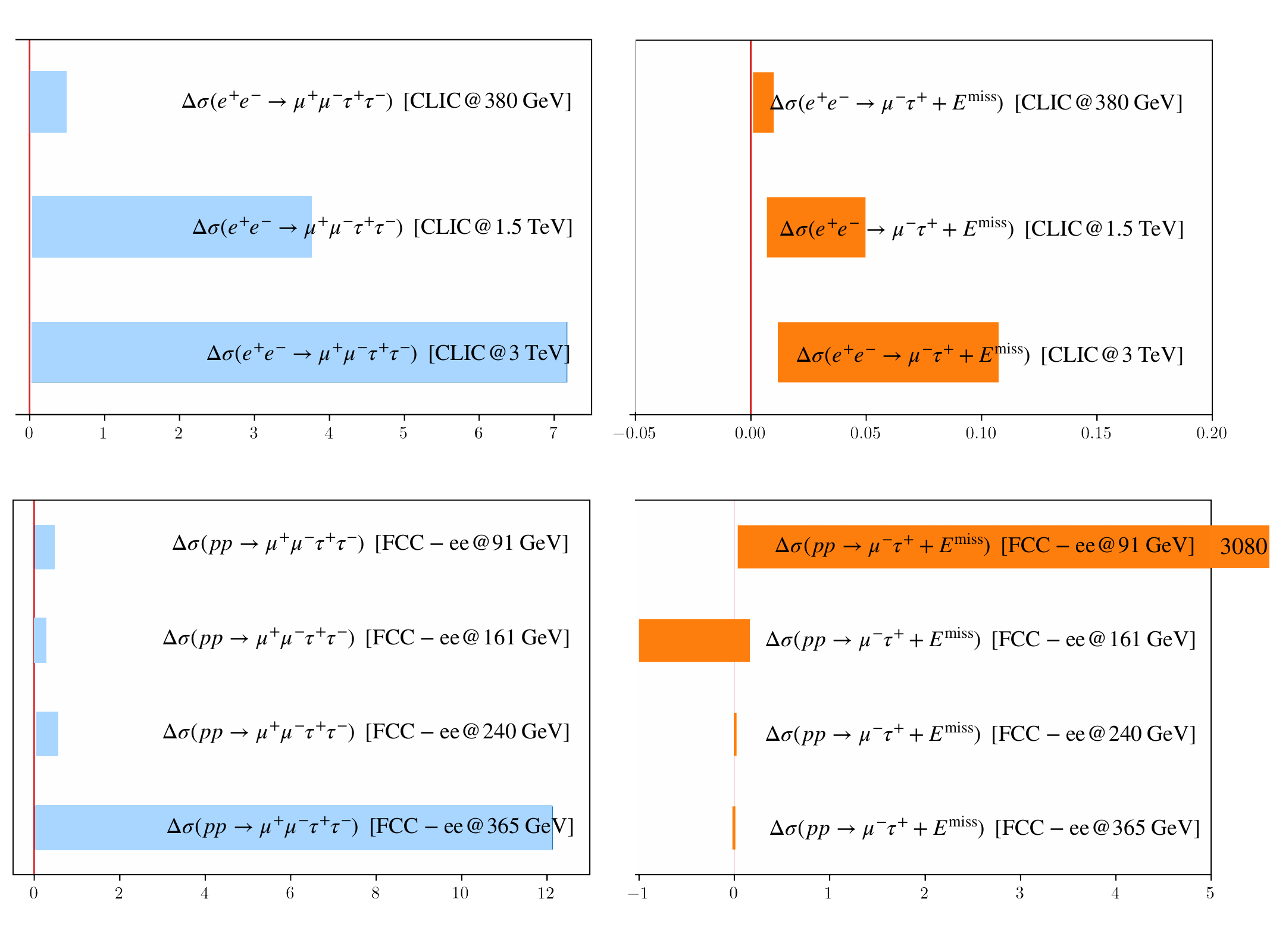}
    \caption{Comparison of the prospects for electron colliders: $\Delta \sigma^{ee}_{Z^\prime \chi}$ (CLIC and FCC-ee at the different $\sqrt s$ runs). The scaling of FCC-ee at the $Z$-pole run has been adapted, as its maximum value is $\Delta \sigma^{ee}_{Z^\prime \chi} \leq 3000$.}
    \label{fig:delta_ee}
\end{figure}

In what follows we limit ourselves to a na\"ive, albeit illustrative comparison: taking only the central values of the predicted cross-sections, we evaluate how distinctive the NP signal can be with respect to the SM background via the ad-hoc quantity
\begin{equation}\label{eq:DeltaSigma:def}
\Delta \sigma^{ff}_{Z^\prime \chi} \,= \,(\sigma_\mathrm{NP}-\sigma_\mathrm{SM})/\sigma_\mathrm{SM}\,,
\end{equation}
in which $ff$ denotes the colliding particles. We recall that $\sigma_\mathrm{NP}$ corresponds to the total cross-section  (i.e. including both the SM and the NP contributions). 
In Figs.~\ref{fig:delta_hadron} -- \ref{fig:delta_ee}, we thus carry out such a comparison for the processes for which it is  possible to do so\footnote{With the exception of $\mu$TRISTAN in its $\mu^+\mu^+$ mode, we notice that processes leading to same-sign final state leptons have no SM background and, thus, cannot be compared using this strategy.}. The spread of the (coloured) horizontal bars corresponds to the variation of the cross-section for all considered $m_{Z^\prime}$ benchmark points, from the smallest to the largest value of $\sigma$.

Although we will not repeat here the previous discussion concerning the prospects for each future collider in detail, this direct graphical comparison succeeds in emphasising the very strong potential of FCC-ee (at the $Z$-pole run) and the unique capabilities of a future muon collider, possibly the most promising machine to unveil the presence of new leptophilic cLFV vector bosons. Indeed, we clearly observe that in these two facilities we may expect deviations reaching factors of a few or, in some cases, orders of magnitude -- both in the case of fully visible final states but also for channels involving missing transverse energy. Hence, we can infer  that both FCC-ee and a future muon collider constitute ideal environments to probe not only the LFV nature of new gauge bosons, but also their capacity to act as portals to dark sectors containing (light, i.e. $m_\chi < m_{Z^\prime}/2$) dark matter candidates. 
It is also interesting to notice that differences with respect to the SM predictions -- reaching several tens of percent in the relevant cross-section  -- could already be present in the available LHC data. 

Concerning processes for which there is no SM contribution, the above comparison relying on $\Delta \sigma^{ff}_{Z^\prime \chi}$ does not apply. However, and from inspection of the distinct facilities, a muon collider (especially in its run at $\sqrt s=1$~TeV) indisputably offers the best prospects, with cross-sections nearing $\mathcal{O}(\text{pb})$.

\section{Conclusions and final discussion}\label{sec:concs}
In this work we have discussed the collider signatures of a minimal SM extension encompassing a strictly cLFV leptophilic 
$Z^\prime$ boson, acting as a dark portal in view of its couplings to a new Majorana stable fermion. Our approach is purely phenomenological (i.e. we refrain from proposing a concrete UV completion or specific $U(1)^\prime$ breaking mechanisms), rather focusing on the potential of such a minimal NP model to address the SM dark matter problem, and reduce the current tension between theory and observation regarding  $(g-2)_\mu$. 

On the dark matter side, the model can reproduce the observed dark matter relic abundance in the Universe via the thermal freeze-out mechanism, while evading constraints stemming from direct and indirect dark matter detection, and CMB measurements -- with the exception of the low  mass regime ($m_\chi \lesssim 5$~GeV) which could be in some tension with the latter two classes of DM searches. Future improvements in both types of measurements could further scrutinise this region of parameter space.

Once the relevant parameters (new couplings and the masses of the NP states) have been constrained, to both saturate $\Delta a_\mu$ and account for a viable DM relic density, we have considered several channels which can be explored at current and future collider facilities, and which could hopefully offer evidence for such a NP model. 

In particular, we have considered the processes 
$f\bar{f} \to \mu^-\mu^+\tau^-\tau^+$, $\bar{f} \to \mu^-\mu^-\tau^+\tau^+$, as well as final states with missing energy (carried by neutrinos and/or the DM candidate)
$f\bar{f} \to \mu^-\tau^+ + E^\mathrm{miss}$, to be searched for at the LHC, CLIC, FCC-ee and FCC-hh, and at a future muon collider (modes exclusive to $\mu$TRISTAN have also been considered). 
Processes leading to same-sign $\mu\tau$ dilepton pairs (i.e. $\mu^\pm\mu^\pm \tau^\mp\tau^\mp$) would offer a true ``smoking gun'' for such a NP realisation. In particular, the associated cross-sections at a future muon collider (running at 1~TeV centre-of-mass energy) are expected to be very large, nearing 0.1~pb (or more). From all the high-energy colliders here investigated, a future muon collider perhaps offers the most promising prospects to probe  the considered NP model: the number of observables to look for is unmatched, and the cross-sections for the 4-lepton final states, as well as for 2-lepton plus missing energy are sizeable (and in general larger than for hadron or $e^+e^-$ machines). Moreover, the associated topology ($t$-channel) opens the door to the identification of clear peaks in the dilepton invariant mass distribution (in association with a resonant $Z^\prime$ production). 
Especially for its run at the $Z$-pole, FCC-ee also offers very promising prospects -- especially for opposite-sign $\mu\tau$ dilepton pairs. 

It is also very relevant to emphasise the strong complementarity of future colliders in probing different $Z^\prime$ mass regimes: while light new bosons (that is $m_{Z^\prime} \leq 50$~GeV) mediate processes which would be more easily ``visible'' at circular hadron and electron colliders (LHC, FCC-ee and FCC-hh), heavier mediators would offer more distinctive signatures (i.e., a clearer separation from SM background  and larger production cross-sections) at future linear and muon colliders.

Decays of the $Z$ boson to doubly flavour-violating 4-lepton final states have been so far the object of few dedicated searches. Such channels ($Z\to \mu^\pm\mu^\pm\tau^\mp \tau^\mp$) would offer excellent grounds to look for NP (including the model here studied). 
At the FCC-ee, running at $\sqrt s=91$~GeV, 
we estimate BR($Z\to \mu^\pm \mu^\pm \tau^\mp \tau^\mp$) to be as large as $\mathcal{O}(5 \times 10^{-6})$
for mediator masses $m_{Z^\prime} < M_Z$. 
Searches for the $Z\to \mu^\pm \mu^\pm \tau^\mp \tau^\mp$ decay would thus offer a complementary probe of this NP model.
We urge experimental collaborations to consider and carry out dedicated searches for these decay modes, in view of the unique potential to unveil NP scenarios. The only relevant signature that appears to exist at the level of LHC searches concerns exotic heavy scalars decaying into opposite-sign dilepton 
pairs~\cite{ATLAS:2023qpu}, although these searches are clearly optimised in view of specific underlying models (considerably different setups to the one here proposed).

Despite the simplicity of the approach here adopted, the results obtained suggest that a minimal SM extension via a cLFV leptophilic $Z^\prime$ portal (and the associated DM candidate) are phenomenologically appealing, and allow to ease SM observational problems and tensions. Moreover, the richness and complementarity of the future collider projects render such an extension testable, and its many peculiar decay modes potentially observable. Once the experimental programmes for the distinct prospective machines are defined, and detector simulations (among others) are available, more comprehensive dedicated studies can then  be carried out.

\section*{Acknowledgements}
This project has received support from the European Union's Horizon 2020 research and innovation programme under the Marie Sk\l{}odowska-Curie grant agreement No.~860881 (HIDDe$\nu$ network) and from the IN2P3 (CNRS) Master Project, ``Flavour probes: lepton sector and beyond'' (16-PH-169).
JK is supported by the Slovenian Research Agency under the research grants N1-0253 and in part by J1-4389. JK is grateful to Luka Leskovec for fruitful discussions.
EP acknowledges financial support from the European Research Council (ERC) under the European Union’s Horizon 2020 research and innovation programme under grant agreement 833280 (FLAY), and from the Swiss National Science Foundation (SNF) under contract 200020-204428.

\appendix
\section{Appendix}\label{sec:app}
\subsection{Relevant bounds from (flavoured) searches: LFUV and cLFV}
\label{app:flavour-bounds}

\paragraph{Lepton flavour universality violation: $Z$ decays}
As mentioned in the main body of the manuscript, 
we consider the ratios of leptonic $Z$ boson decay widths 
\begin{equation}
    R^Z_{\alpha\beta} \,= \,\dfrac{\Gamma(Z \to \ell_\alpha^+\ell_\alpha^-)}{\Gamma(Z \to \ell_\beta^+\ell_\beta^-)}\,, \quad \text{with } \alpha \neq \beta \, =\, e, \, \mu, \, \tau\,.
\end{equation}
Let us recall  that the SM predictions for these ratios (at 2-loop accuracy) are~\cite{Freitas:2014hra}
\begin{equation}
    {R^Z_{\mu e}}|_\text{SM} \,=\,1\,, \quad\quad {R^Z_{\tau \mu}}|_\text{SM} \,=\,0.9977\,, \quad\quad {R^Z_{\tau e}}|_\text{SM} \,=\,0.9977\,,
\end{equation}
with negligible associated uncertainties. The above theoretical predictions  must be compared with the corresponding experimental values~\cite{ParticleDataGroup:2020ssz},
\begin{eqnarray}
   {R^Z_{\mu e}}|_\text{exp} &=& 1.0001 \pm 0.0024 \,, \quad\quad
    {R^Z_{\tau \mu}}|_\text{exp} = 1.0010 \pm 0.0026\,,\nonumber \\
     {R^Z_{\tau e}}|_\text{exp} &=& 1.0020 \pm 0.0032\,.
\label{eq::ZdecayLFU}
\end{eqnarray} 

\paragraph{Lepton flavour universality violation: $\tau$ decays}
Further probes of LFU in lepton decays can be constructed from comparing the leptonic  decay widths of the tau-lepton, 
\begin{equation}
        R^\tau_{\mu e} \,\equiv\, \dfrac{\Gamma(\tau^- \to \mu^- \nu_\tau \overline{\nu}_\mu)}{\Gamma(\tau^- \to e^- \nu_\tau \overline{\nu}_e)} \,,
\label{eq:def:tauratio}
\end{equation}
whose SM prediction is ${R^\tau_{\mu e}}|_\text{SM} = 0.972564 \pm 0.00001$~\cite{Pich:2009zza}. The latter is in goood agreement with the experimental measurements\footnote{We notice  that experimentally one cannot tag individual neutrino flavours, so that generic decays $\tau \to \ell \nu_\gamma \overline{\nu}_\delta$, with $\gamma,\delta = e, \mu, \tau$ contribute to 
${R^\tau_{\mu e}}|_\text{exp}$. } (from ARGUS~\cite{ARGUS:1991zhv}, CLEO~\cite{CLEO:1996oro} and BaBar~\cite{BaBar:2009lyd}), as reported by the HFLAV collaboration global fit~\cite{HFLAV:2019otj}:
\begin{equation}
        {R^\tau_{\mu e}}|_\text{exp}\,\, \equiv\, \,\dfrac{\Gamma(\tau^- \to \mu^- \nu_\tau \overline{\nu}_\mu)}{\Gamma(\tau^- \to e^- \nu_\tau \overline{\nu}_e)}\,=\,0.9761\pm 0.0028\,.
\label{eq::tauratio}
\end{equation}

\paragraph{Charged lepton flavour violating  processes}
Rare cLFV transitions and decays offer powerful probes 
of NP contributions stemming from the presence 
of an LFV leptophilic $Z^\prime$.
The most relevant processes include 
cLFV muon and tau three-body decays ($\mu \to 3e$, $\tau \to 3e$, $\tau \to 3\mu$, $\tau^- \to \mu^- e^+ e^-$, $\tau^- \to \mu^- e^+ \mu^-$, $\tau^- \to e^- \mu^+ \mu^-$ and $\tau^- \to e^- \mu^+ e^-$),  and radiative decays ($\mu \to e \gamma$, $\tau \to e \gamma$ and $\tau \to \mu \gamma$), as well as neutrinoless conversion in nuclei. The experimental bounds (and  future sensitivities) for these decays are collected  in Table~\ref{tab:cLFVdata}.
\renewcommand{\arraystretch}{1.3}
\begin{table}[h!]
    \centering
    \hspace*{-7mm}{\small\begin{tabular}{|c|c|c|}
    \hline
    Observable & Current bound & Future sensitivity  \\
    \hline\hline
    $\text{BR}(\mu\to e \gamma)$    &
    \quad $<3.1\times 10^{-13}$ \quad (MEG~\cite{MEGII:2023ltw})   &
    \quad $6\times 10^{-14}$ \quad (MEG II~\cite{Baldini:2018nnn}) \\
    $\text{BR}(\tau \to e \gamma)$  &
    \quad $<3.3\times 10^{-8}$ \quad (BaBar~\cite{Aubert:2009ag})    &
    \quad $3\times10^{-9}$ \quad (Belle II~\cite{Kou:2018nap})      \\
    $\text{BR}(\tau \to \mu \gamma)$    &
     \quad $ <4.4\times 10^{-8}$ \quad (BaBar~\cite{Aubert:2009ag})  &
    \quad $10^{-9}$ \quad (Belle II~\cite{Kou:2018nap})     \\
    \hline
    $\text{BR}(\mu \to 3 e)$    &
     \quad $<1.0\times 10^{-12}$ \quad (SINDRUM~\cite{Bellgardt:1987du})    &
     \quad $10^{-15(-16)}$ \quad (Mu3e~\cite{Blondel:2013ia})   \\
    $\text{BR}(\tau \to 3 e)$   &
    \quad $<2.7\times 10^{-8}$ \quad (Belle~\cite{Hayasaka:2010np})&
    \quad $5\times10^{-10}$ \quad (Belle II~\cite{Kou:2018nap})     \\
    $\text{BR}(\tau \to 3 \mu )$    &
    \quad $<3.3\times 10^{-8}$ \quad (Belle~\cite{Hayasaka:2010np})  &
    \quad $5\times10^{-10}$ \quad (Belle II~\cite{Kou:2018nap})     \\
    & & \quad$5\times 10^{-11}$\quad (FCC-ee~\cite{Abada:2019lih})\\
        $\text{BR}(\tau^- \to e^-\mu^+\mu^-)$   &
    \quad $<2.7\times 10^{-8}$ \quad (Belle~\cite{Hayasaka:2010np})&
    \quad $5\times10^{-10}$ \quad (Belle II~\cite{Kou:2018nap})     \\
    $\text{BR}(\tau^- \to \mu^-e^+e^-)$ &
    \quad $<1.8\times 10^{-8}$ \quad (Belle~\cite{Hayasaka:2010np})&
    \quad $5\times10^{-10}$ \quad (Belle II~\cite{Kou:2018nap})     \\
    $\text{BR}(\tau^- \to e^-\mu^+e^-)$ &
    \quad $<1.5\times 10^{-8}$ \quad (Belle~\cite{Hayasaka:2010np})&
    \quad $3\times10^{-10}$ \quad (Belle II~\cite{Kou:2018nap})     \\
    $\text{BR}(\tau^- \to \mu^-e^+\mu^-)$   &
    \quad $<1.7\times 10^{-8}$ \quad (Belle~\cite{Hayasaka:2010np})&
    \quad $4\times10^{-10}$ \quad (Belle II~\cite{Kou:2018nap})     \\
    \hline
    $\text{CR}(\mu- e, \text{N})$ &
     \quad $<7 \times 10^{-13}$ \quad  (Au, SINDRUM~\cite{Bertl:2006up}) &
    \quad $10^{-14}$  \quad (SiC, DeeMe~\cite{Nguyen:2015vkk})    \\
    & &  \quad $2.6\times 10^{-17}$  \quad (Al, COMET~\cite{Krikler:2015msn,Adamov:2018vin})  \\
    & &  \quad $8 \times 10^{-17}$  \quad (Al, Mu2e~\cite{Bartoszek:2014mya})\\
    \hline
    \hline 
    $\mathrm{BR}(Z\to e^\pm\mu^\mp)$ & \quad$< 4.2\times 10^{-7}$\quad (ATLAS~\cite{Aad:2014bca}) & \quad$\mathcal O (10^{-10})$\quad (FCC-ee~\cite{Abada:2019lih})\\
    $\mathrm{BR}(Z\to e^\pm\tau^\mp)$ & \quad$< 5.2\times 10^{-6}$\quad (OPAL~\cite{Akers:1995gz}) & \quad$\mathcal O (10^{-10})$\quad (FCC-ee~\cite{Abada:2019lih})\\
    $\mathrm{BR}(Z\to \mu^\pm\tau^\mp)$ & \quad$< 5.4\times 10^{-6}$\quad (OPAL~\cite{Akers:1995gz}) & \quad $\mathcal O (10^{-10})$\quad (FCC-ee~\cite{Abada:2019lih})\\
    \hline
    \end{tabular}}
    \caption{Current experimental bounds and future sensitivities on cLFV observables considered in this work. All limits are given at $90\%\:\mathrm{C.L.}$ (notice that the Belle II sensitivities correspond to an integrated luminosity of $50\:\mathrm{ab}^{-1}$). }
    \label{tab:cLFVdata}
\end{table}
\renewcommand{\arraystretch}{1.}

Muonium oscillations can  also lead to stringent constraints on $Z^\prime$ extensions of the  SM, as the latter can be at the source of new contributions to the transition probability $P$; the theoretical predictions should be compared with  
$  P\,< \,8.3 \times 10^{-11}$, 
as set by the PSI experiment~\cite{Willmann:1998gd}.

\subsection{Feynman diagrams}
\label{app:feynman}
In this Appendix we collect Feynman diagrams which illustrate the topologies of the processes relevant to the discussion of Section~\ref{sec:colliders}, in particular concerning muon colliders.

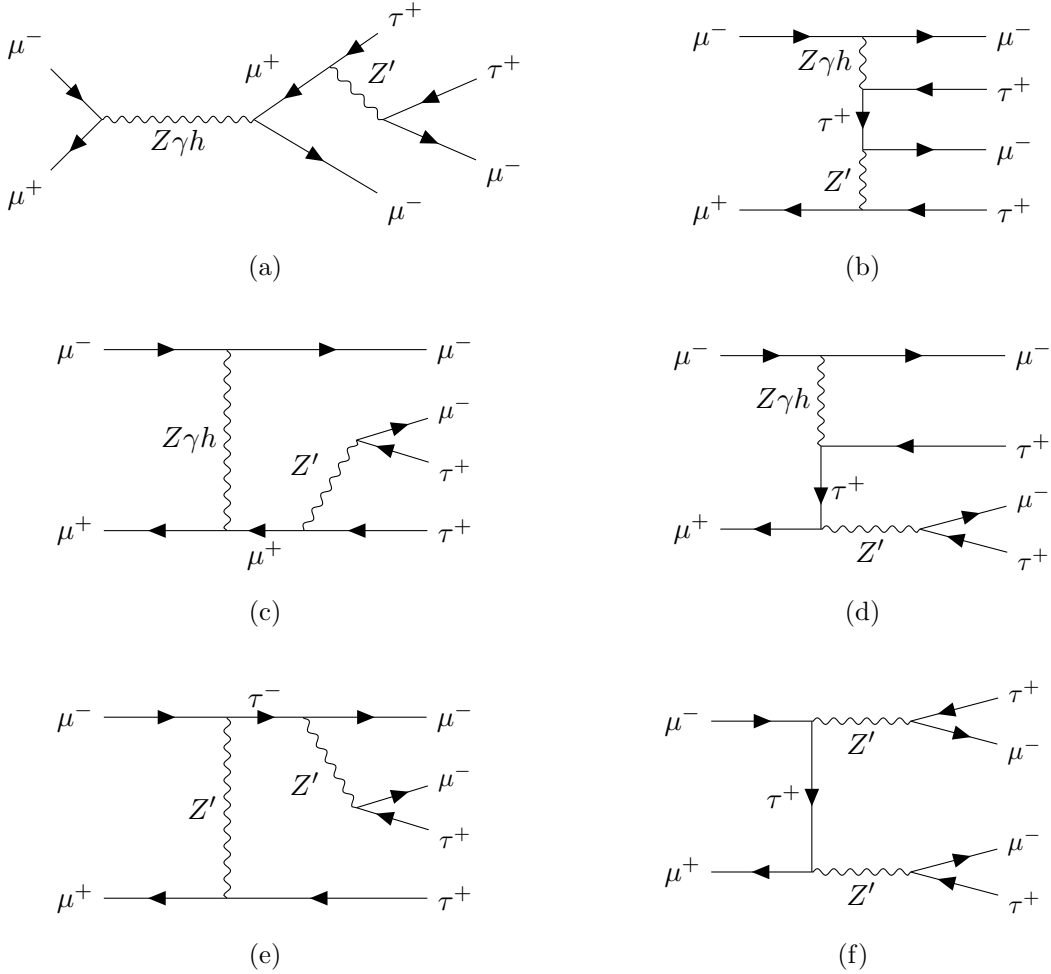
\begin{figure}[h!]
    \centering
\begin{subfigure}[b]{0.48\textwidth}\centering  \begin{tikzpicture}
    \begin{feynman}
    \vertex (q1) at (-1,1) {\(\mu^-\)};
    \vertex (q2) at (-1,-1){\(\mu^+\)};
    \vertex (a) at (-0 ,0) ;
    \vertex (b) at (2,0);
    \vertex (c) at (4,1.4) {\(\tau^+\)};
    \vertex (d) at (4,-1.2){\(\mu^-\)}; 
    \vertex (e) at (3,0.7);
    \vertex (f) at (3.7,0);
    \vertex (f1) at (5.3,0.7){\(\tau^+  \)};
    \vertex (f2) at (5.3,-0.7){\(\mu^- \)};
    \diagram* {
    (q1) -- [fermion] (a) -- [fermion] (q2),
    (a) -- [boson, edge label'=\(Z\gamma h\)] (b),
    (c) -- [fermion] (e) -- [fermion, edge label'=\(\mu^+\)] (b) -- [fermion] (d),
    (e) -- [boson, edge label=\(Z^\prime\)] (f),
    (f1) -- [fermion](f) -- [fermion](f2)};
    \end{feynman}
\end{tikzpicture}\caption*{(a)}
            \label{}
    \end{subfigure}
    \hfill
\begin{subfigure}[b]{0.48\textwidth}\centering  \begin{tikzpicture}
    \begin{feynman}
    \vertex (mu1) at (0,1.5) {\(\mu^-\)};
    \vertex (mu2) at (0,-0.8){\( \mu^+\)};
    \vertex (a) at (2, 1.5);
    \vertex (b) at (2, 0.8);
    \vertex (c) at (2, 0);
    \vertex (d) at (2, -0.8);
    \vertex (f1) at (4.,1.5) {\(\mu^-\)};
    \vertex (f2) at (4.,-0.8){\( \tau^+\)};
    \vertex (m1) at (4.,0) {\(\mu^-\)};
    \vertex (m2) at (4.,0.8){\( \tau^+\)};
    \diagram* {(mu1) -- [fermion] (a)  -- [fermion] (f1),
    (a) -- [boson, edge label'=\(Z \gamma h\)] (b),
    (m2) -- [fermion] (b)-- [fermion, edge label'=\(\tau^+\)] (c)-- [fermion] (m1),
    (c) -- [boson, edge label'=\(Z^\prime\)] (d),
    (mu2) -- [anti fermion] (d)  -- [anti fermion] (f2),
    };
    \end{feynman}
\end{tikzpicture}\caption*{(b)}
            \label{}
    \end{subfigure}
\\ \vspace{5mm}
\begin{subfigure}[b]{0.48\textwidth}\centering\begin{tikzpicture}
    \begin{feynman}
    \vertex (mu1) at (0,1.2) {\(\mu^-\)};
    \vertex (mu2) at (0,-1.2){\( \mu^+\)};
    \vertex (a) at (2,1.2) ;
    \vertex (b) at (2,-1.2);
    \vertex (c) at (5,1.2) {\(\mu^-\)};
    \vertex (d) at (5,-1.2){\(\tau^+\)}; 
    \vertex (e) at (3,-1.2);
    \vertex (f) at (3.7,0);
    \vertex (f1) at (5,0.4){\small\(\mu^- \)};
    \vertex (f2) at (5,-0.4){\small\(\tau^+ \)};
    \diagram* {
    (mu1) -- [fermion] (a)  -- [fermion] (c),
    (a) -- [boson, edge label'=\(Z \gamma h\)] (b),
    (mu2) -- [anti fermion] (b)-- [anti fermion, edge label'=\(\mu^+\)] (e) -- [anti fermion] (d),
    (e) -- [boson, edge label=\(Z^\prime\)] (f),
    (f1) -- [anti fermion](f) -- [anti fermion](f2)};
    \end{feynman}
\end{tikzpicture}\caption*{(c)}
            \label{}
    \end{subfigure}
    \hfill
\begin{subfigure}[b]{0.48\textwidth}\centering\begin{tikzpicture}
    \begin{feynman}
    \vertex (mu1) at (0,1.5) {\(\mu^-\)};
    \vertex (mu2) at (0,-0.8){\( \mu^+\)};
    \vertex (a) at (1.7,1.5) ;
    \vertex (ab) at (1.7,0.3) ;
    \vertex (b) at (1.7,-0.8);
    \vertex (c) at (4.5,1.5) {\(\mu^-\)};
    \vertex (d) at (3,-0.8); 
    \vertex (fm2) at (4.5,0.3) {\(\tau^+\)};
    \vertex (f1a) at (4.5,-0.4){\small\(\mu^- \)};
    \vertex (f2a) at (4.5,-1.2){\small\(\tau^+ \)};
    \diagram* {
    (mu1) -- [fermion] (a) -- [fermion] (c),
    (a) -- [boson, edge label'=\(Z \gamma h\)] (ab),
    (b) -- [boson, edge label'=\(Z^\prime\)] (d),
    (f1a) -- [anti fermion](d) -- [anti fermion](f2a),
    (mu2) -- [anti fermion] (b) -- [anti fermion, edge label'=\(\tau^+\)] (ab) -- [anti fermion] (fm2)};
    \end{feynman}
\end{tikzpicture}\caption*{(d)}
            \label{}
    \end{subfigure}
\\ \vspace{5mm}
\begin{subfigure}[b]{0.48\textwidth}\centering\begin{tikzpicture}
    \begin{feynman}
    \vertex (mu1) at (0,1.2) {\(\mu^-\)};
    \vertex (mu2) at (0,-1.2){\( \mu^+\)};
    \vertex (a) at (2,1.2) ;
    \vertex (b) at (2,-1.2);
    \vertex (c) at (5,1.2) {\(\mu^-\)};
    \vertex (d) at (5,-1.2){\(\tau^+\)}; 
    \vertex (e) at (3,1.2);
    \vertex (f) at (3.7,0);
    \vertex (f1) at (5,0.4){\small\(\mu^- \)};
    \vertex (f2) at (5,-0.4){\small\(\tau^+ \)};
    \diagram* {
    (mu1) -- [fermion] (a) -- [fermion, edge label=\(\tau^-\)] (e) -- [fermion] (c),
    (a) -- [boson, edge label'=\(Z^\prime\)] (b),
    (mu2) -- [anti fermion] (b) -- [anti fermion] (d),
    (e) -- [boson, edge label'=\(Z^\prime\)] (f),
    (f1) -- [anti fermion](f) -- [anti fermion](f2)};
    \end{feynman}
\end{tikzpicture}\caption*{(e)}
            \label{}
    \end{subfigure}
    \hfill
\begin{subfigure}[b]{0.48\textwidth}\centering\begin{tikzpicture}
    \begin{feynman}
    \vertex (mu1) at (0,1.2) {\(\mu^-\)};
    \vertex (mu2) at (0,-0.8){\( \mu^+\)};
    \vertex (a) at (1.7,1.2) ;
    \vertex (b) at (1.7,-0.8);
    \vertex (c) at (3,1.2) ;
    \vertex (d) at (3,-0.8); 
    \vertex (f1) at (4.5,0.8){\small\(\mu^- \)};
    \vertex (f2) at (4.5,1.6){\small\(\tau^+ \)};
    \vertex (f1a) at (4.5,-0.4){\small\(\mu^- \)};
    \vertex (f2a) at (4.5,-1.2){\small\(\tau^+ \)};
    \diagram* {
    (mu1) -- [fermion] (a) -- [fermion, edge label'=\(\tau^+\)] (b) -- [fermion] (mu2),
    (a) -- [boson, edge label'=\(Z^\prime\)] (c),
    (b) -- [boson, edge label'=\(Z^\prime\)] (d),
    (f1a) -- [anti fermion](d) -- [anti fermion](f2a),
    (f1) -- [anti fermion](c) -- [anti fermion](f2)};
    \end{feynman}
\end{tikzpicture}\caption*{(f)}
            \label{}
    \end{subfigure}
    \hfill
    \caption{Different topologies arising in the NP model here considered, all contributing to $\mu^+\mu^-\to \mu^-\mu^-\tau^+\tau^+$.}
    \label{fig:diag:mucoll_ssl}
\end{figure}

\paragraph{Muon collider}
In Fig.~\ref{fig:diag:mucoll_ssl} we display the relevant Feynman diagrams with a $Z^\prime$ contributing to $\mu^+\mu^-\to \mu^-\mu^-\tau^+\tau^+$ at a future muon collider. For simplicity, we do not display the SM only mediated diagrams.
As noticed, the present set is only illustrative; additional diagrams can be obtained, for instance, by exchanging $\mu-\tau$.

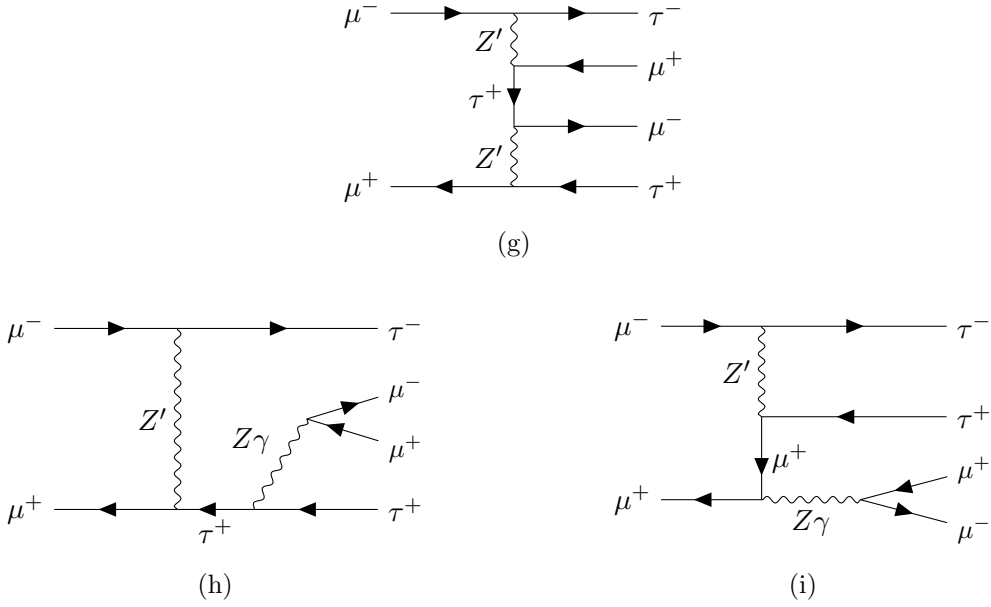
\begin{figure}[h!]
    \centering
\begin{subfigure}[b]{0.48\textwidth}\centering  \begin{tikzpicture}
    \begin{feynman}
    \vertex (mu1) at (0,1.5) {\(\mu^-\)};
    \vertex (mu2) at (0,-0.8){\( \mu^+\)};
    \vertex (a) at (2, 1.5);
    \vertex (b) at (2, 0.8);
    \vertex (c) at (2, 0);
    \vertex (d) at (2, -0.8);
    \vertex (f1) at (4.,1.5) {\(\tau^-\)};
    \vertex (f2) at (4.,-0.8){\( \tau^+\)};
    \vertex (m1) at (4.,0) {\(\mu^-\)};
    \vertex (m2) at (4.,0.8){\( \mu^+\)};
    \diagram* {(mu1) -- [fermion] (a)  -- [fermion] (f1),
    (a) -- [boson, edge label'=\(Z^\prime\)] (b),
    (m2) -- [fermion] (b)-- [fermion, edge label'=\(\tau^+\)] (c)-- [fermion] (m1),
    (c) -- [boson, edge label'=\(Z^\prime\)] (d),
    (mu2) -- [anti fermion] (d)  -- [anti fermion] (f2),
    };
    \end{feynman}
\end{tikzpicture}\caption*{(g)}
            \label{}
    \end{subfigure}
\\ \vspace{5mm}
\begin{subfigure}[b]{0.48\textwidth}\centering  \begin{tikzpicture}
    \begin{feynman}
    \vertex (mu1) at (0,1.2) {\(\mu^-\)};
    \vertex (mu2) at (0,-1.2){\( \mu^+\)};
    \vertex (a) at (2,1.2) ;
    \vertex (b) at (2,-1.2);
    \vertex (c) at (5,1.2) {\(\tau^-\)};
    \vertex (d) at (5,-1.2){\(\tau^+\)}; 
    \vertex (e) at (3,-1.2);
    \vertex (f) at (3.7,0);
    \vertex (f1) at (5,0.4){\small\(\mu^- \)};
    \vertex (f2) at (5,-0.4){\small\(\mu^+ \)};
    \diagram* {
    (mu1) -- [fermion] (a)  -- [fermion] (c),
    (a) -- [boson, edge label'=\(Z^\prime\)] (b),
    (mu2) -- [anti fermion] (b)-- [anti fermion, edge label'=\(\tau^+\)] (e) -- [anti fermion] (d),
    (e) -- [boson, edge label=\(Z \gamma\)] (f),
    (f1) -- [anti fermion](f) -- [anti fermion](f2)};
    \end{feynman}
\end{tikzpicture}\caption*{(h)}
            \label{}
    \end{subfigure}
    \hfill
\begin{subfigure}[b]{0.48\textwidth}\centering  \begin{tikzpicture}
    \begin{feynman}
    \vertex (mu1) at (0,1.5) {\(\mu^-\)};
    \vertex (mu2) at (0,-0.8){\( \mu^+\)};
    \vertex (a) at (1.7,1.5) ;
    \vertex (ab) at (1.7,0.3) ;
    \vertex (b) at (1.7,-0.8);
    \vertex (c) at (4.5,1.5) {\(\tau^-\)};
    \vertex (d) at (3,-0.8); 
    \vertex (fm2) at (4.5,0.3) {\(\tau^+\)};
    \vertex (f1a) at (4.5,-0.4){\small\(\mu^+ \)};
    \vertex (f2a) at (4.5,-1.2){\small\(\mu^- \)};
    \diagram* {
    (mu1) -- [fermion] (a) -- [fermion] (c),
    (a) -- [boson, edge label'=\(Z^\prime \)] (ab),
    (b) -- [boson, edge label'=\(Z \gamma\)] (d),
    (f1a) -- [fermion](d) -- [fermion](f2a),
    (mu2) -- [anti fermion] (b) -- [anti fermion, edge label'=\(\mu^+\)] (ab) -- [anti fermion] (fm2)};
    \end{feynman}
\end{tikzpicture}\caption*{(i)}
            \label{}
    \end{subfigure}
    \hfill
    \caption{Illustration of the different topologies of the processes contributing to $\mu^+\mu^-\to \mu^-\mu^+\tau^-\tau^+$.}
    \label{fig:diag:mucoll_osl}
\end{figure}

Considering the opposite-sign dilepton pair in the final state, the NP Feynman diagrams can be obtained from the ones collected in Fig.~\ref{fig:diag:mucoll_ssl} (with the appropriate replacements of lepton charges), together with the diagrams in Fig.\ref{fig:diag:mucoll_osl}.

For the case of missing energy in the final state (neutrinos and DM), all previous diagrams  from Fig.~\ref{fig:diag:mucoll_ssl} and Fig.~\ref{fig:diag:mucoll_osl} (with the exception of the first one (top) in Fig.~\ref{fig:diag:mucoll_osl}) do contribute, under the appropriate replacements -- in particular $W^\pm$ mediated contributions (leading to $\ell^\pm \nu$).

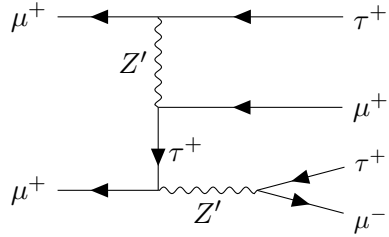
\begin{figure}[h!]
    \centering
\begin{tikzpicture}
    \begin{feynman}
    \vertex (mu1) at (0,1.5) {\(\mu^+\)};
    \vertex (mu2) at (0,-0.8){\( \mu^+\)};
    \vertex (a) at (1.7,1.5) ;
    \vertex (ab) at (1.7,0.3) ;
    \vertex (b) at (1.7,-0.8);
    \vertex (c) at (4.5,1.5) {\(\tau^+\)};
    \vertex (d) at (3,-0.8); 
    \vertex (fm2) at (4.5,0.3) {\(\mu^+\)};
    \vertex (f1a) at (4.5,-0.4){\small\(\tau^+ \)};
    \vertex (f2a) at (4.5,-1.2){\small\(\mu^- \)};
    \diagram* {
    (mu1) -- [anti fermion] (a) -- [anti fermion] (c),
    (a) -- [boson, edge label'=\(Z^\prime \)] (ab),
    (b) -- [boson, edge label'=\(Z^\prime\)] (d),
    (f1a) -- [fermion](d) -- [fermion](f2a),
    (mu2) -- [anti fermion] (b) -- [anti fermion, edge label'=\(\tau^+\)] (ab) -- [anti fermion] (fm2)};
    \end{feynman}
\end{tikzpicture}
    \caption{Further contributing topologies to  $\mu^+\mu^+\to \mu^-\mu^+\tau^+\tau^+$ (at $\mu$TRISTAN).}
    \label{fig:diag:muT:osmu}
\end{figure}

\paragraph{$\mu$TRISTAN}
The new contributions to the process $\mu^+\mu^+\to \mu^-\mu^+\tau^+\tau^+$ at $\mu$TRISTAN arise from the topologies depicted in Figs.~\ref{fig:diag:mucoll_ssl} and~\ref{fig:diag:mucoll_osl} (with the appropriate replacements, and with the exception of the first and last diagrams of Fig.~\ref{fig:diag:mucoll_ssl}). Moreover, one also has the contribution displayed in Fig.~\ref{fig:diag:muT:osmu}.

At $\mu$TRISTAN, the contributions to the process with opposite-sign tau pair $\mu^+\mu^+\to \mu^+\mu^+\tau^-\tau^+$ arise from the Feynman diagrams depicted in Fig.~\ref{fig:diag:mucoll_ssl} (with the exception of diagrams (a) and (f), as they cannot appear in a $\mu^+\mu^+$ collider) together with the diagram in Fig.~\ref{fig:diag:muT:osmu}. 
The same contributions apply to the process with missing energy $\mu^+\mu^+\to \mu^+\tau^++ E^\mathrm{miss}$,  under the appropriate replacements -- in particular $W^\pm$ mediated contributions leading to neutrinos in the final state.
Finally when considering $\mu$TRISTAN in $e^-\mu^+$ collision mode, the processes $e^-\mu^+\to e^-\mu^-\tau^+\tau^+$ and $e^-\mu^+\to e^-\mu^+\tau^+\tau^-$ receive NP contributions mediated by the diagrams (b), (c) and (d) from Fig.~\ref{fig:diag:mucoll_ssl} (replacing $\mu^-$ by $e^-$). 
Notice that $e^-\mu^+\to e^-\mu^+\tau^+\tau^-$ also receives pure SM contributions that are not displayed here.
Moreover the process $e^-\mu^+\to e^-\tau^+ + E^\mathrm{miss}$ is also mediated by diagrams (b), (c) and (d) of Fig.~\ref{fig:diag:mucoll_ssl}.

\newpage

\bibliographystyle{JHEP}
\bibliography{GKPT-references}
\end{document}